\documentclass[12pt,letterpaper]{JHEP3}
\usepackage{amsmath,amssymb,amsfonts,xspace}
\usepackage{epsfig}
\usepackage{epstopdf}

\def\lfig#1#2#3#4#5{
 \begin{figure}
 \refstepcounter{figure}
 \label{#4}
 \addtocounter{figure}{-1}
 \epsfxsize=#3
 \centerline{\epsfbox{#2}}
 \vspace{#5}
 {\bf \caption{{\rm #1}}}
 \end{figure}
}

\numberwithin{equation}{section}

\def\varpi{t}

\def\Im{\,{\rm Im}\,}
\def\Re{\,{\rm Re}\,}

\def\({\left(}
\def\){\right)}
\def\[{\left[}
\def\]{\right]}
\def\<{\left\langle}
\def\>{\right\rangle}
\def\hf{{1\over 2}}

\newcommand{\de}{\mathrm{d}}

\newcommand{\I}{\mathrm{i}}
\newcommand{\e}{\mathrm{e}}

\newcommand{\p}{\partial}

\newcommand{\half}{\frac{1}{2}}

\newcommand{\cV}{\mathcal{V}}

\newcommand{\cS}{\mathcal{S}}

\newcommand{\cK}{\mathcal{K}}
\newcommand{\cM}{\mathcal{M}}

\newcommand{\cN}{\mathcal{N}}

\newcommand{\cX}{\mathcal{X}}

\newcommand{\cR}{\mathcal{R}}

\DeclareSymbolFont{AMSa}{U}{msa}{m}{n}
\DeclareSymbolFont{AMSb}{U}{msb}{m}{n}
\DeclareMathSymbol{\fieldR}{\mathalpha}{AMSb}{"52}

\newcommand{\N}{{\mathcal N}}
\newcommand{\kahler}{{K\"ahler}\xspace}
\newcommand{\hk}{{hyperk\"ahler}\xspace}
\newcommand{\qk}{{quaternion-K\"ahler}\xspace}

\newcommand{\R}{{\mathbb R}}
\newcommand{\Z}{{\mathbb Z}}
\newcommand{\cZ}{\mathcal{Z}}

\newcommand{\cO}{\mathcal{O}}
\newcommand{\cH}{\mathcal{H}}
\newcommand{\cU}{\mathcal{U}}

\newcommand{\pa}{\partial}
\newcommand{\nn}{\nonumber}

\newcommand{\eps}{\epsilon}
\newcommand{\IR}{\mathbb{R}}
\newcommand{\IC}{\mathbb{C}}
\newcommand{\IZ}{\mathbb{Z}}

\newcommand{\sgn}{\mbox{sgn}}
\newcommand{\tzeta}{\tilde\zeta}
\newcommand{\tsigma}{\tilde\sigma}

\newcommand{\txi}{\tilde\xi}

\newcommand{\CP}{\IC P^1}

\def\bea{\begin{eqnarray}}
\def\eea{\end{eqnarray}}
\def\be{\begin{equation}}
\def\ee{\end{equation}}
\def\ba{\begin{align}}
\def\ea{\end{align}}
\def\bse{\begin{subequations}}
\def\ese{\end{subequations}}

\def\ba{\bar a}

\def\bz{\bar z}
\def\bY{\bar Y}

\def\bG{ \bar G }

\def\bF{\bar F}

\def\ui#1{^{[#1]}}

\def\txii#1{{\tilde\xi}^{[#1]}}

\def\ai#1{{\alpha}^{[#1]}}

\def\xii#1{\xi_{[#1]}}

\def\Hij#1{H^{[#1]}}

\newcommand{\Li}{{\rm Li}}

\def\Xikl{\Xi_{\gamma}}

\def\XXint#1#2#3{{\setbox0=\hbox{$#1{#2#3}{\int}$}
\vcenter{\hbox{$#2#3$}}\kern-.5\wd0}}

\def\hHij#1{H^{[#1]}}

\newcommand{\hCX}{\mathcal{X}}

\def\cij#1{c}
\def\ci#1{c}

\def\hnkl{\Omega(\gamma)}

\def\lvol{{\rm cl}}

\def\ws{{\rm w.s.}}

\def\CY{\mathfrak{Y}}
\def\CYm{\mathfrak{\hat Y}}

\def\KC{\cK_C}
\def\KK{\cK_K}

\def\gsl{g}
\def\Fcl{F^{\rm cl}}

\def\bFcl{\bF^{\rm cl}}

\def\varpiqk{t}
\def\hcU{{\cal U}}
\def\qHij#1{H^{[#1]}}
\newcommand{\beq}{\begin{eqnarray}}
\newcommand{\eeq}{\end{eqnarray}}
\def\tleta{\tilde\eta}
\def\tzp{\varpi_+^{c,d}}
\def\tzm{\varpi_-^{c,d}}
\def\tzpm{\varpi_\pm^{c,d}}

\def\tzpcd#1{\varpi_+^{#1}}
\def\tzmcd#1{\varpi_-^{#1}}
\def\tzpmcd#1{\varpi_\pm^{#1}}

\def\cla{\tilde c_a}
\def\cl0{\tilde c_0}

\title{S-duality in Twistor Space}

\preprint{L2C:12-064 \\ CERN-PH-TH/2012-156}

\author{
Sergei Alexandrov$^{1,2}$, Boris Pioline$^{3,4}$
\\
$^1$ {\it Universit\'e Montpellier 2, Laboratoire Charles Coulomb UMR 5221, F-34095,
Montpellier, France}\\

$^2$ {\it CNRS, Laboratoire Charles Coulomb UMR 5221, F-34095,
Montpellier, France}\\

$^3$ {\it CERN PH-TH,
Case C01600, CERN, CH-1211 Geneva 23, Switzerland}\\

$^4$ {\it Laboratoire de Physique Th\'eorique et Hautes
Energies, CNRS UMR 7589, \\
Universit\'e Pierre et Marie Curie,
4 place Jussieu, 75252 Paris cedex 05, France} \\

\vspace*{2mm} {\tt e-mail:
\email{salexand@univ-montp2.fr},
\email{boris.pioline@cern.ch}
}

\vspace*{-3mm}

}

\abstract{In type IIB string compactifications on a Calabi-Yau threefold,
the hypermultiplet moduli space $\cM_H$ must carry an isometric action of the modular group $SL(2,\IZ)$,
inherited from the S-duality symmetry of type IIB string theory in ten dimensions.
We investigate how this modular symmetry is realized at the level of the twistor space of $\cM_H$,
and construct a general class of
$SL(2,\IZ)$-invariant quaternion-K\"ahler metrics with two commuting isometries,
parametrized by a suitably covariant family of holomorphic transition functions.
This family should include $\cM_H$ corrected by D3-D1-D(-1)-instantons
(with fivebrane corrections ignored) and, after taking a suitable rigid limit,
the Coulomb branch of five-dimensional $\cN=2$ gauge theories compactified
on a torus, including monopole string instantons.
These results allow us to considerably simplify the derivation of the mirror map
between type IIA and IIB fields in the sector where only D1-D(-1)-instantons are retained.
}

\begin{document}

\section{Introduction}

String  vacua with $\cN=2$ supersymmetry in four dimensions offer a fascinating vantage point
on the non-perturbative spectrum and symmetries of string theory.
In the low energy approximation, the dynamics around these vacua is described by
the metric on the moduli space parametrized by the massless scalar fields arising
after compactification from ten to four dimensions.
This space factorizes into the direct product of the vector multiplet (VM) moduli space
$\cM_V$ and the hypermultiplet (HM) moduli space $\cM_H$ \cite{deWit:1984px}, which
encode distinct properties of the internal manifold. Whereas the
geometry of $\cM_V$ has been under control for many years, a complete description
of the geometry of $\cM_H$ at the non-perturbative level is still missing. This
is in part due to the fact that the metric on $\cM_H$ must be \qk (QK) \cite{Bagger:1983tt},
a property which is far more complicated to enforce than
special K\"ahler geometry in the vector multiplet sector.

Nevertheless, remarkable progress was achieved in recent years in
the context of type II strings compactified on a Calabi-Yau (CY) threefold,
based on the development of twistorial methods for \qk geometry (see
\cite{Alexandrov:2011va} for an extensive review). This approach was developed in the
mathematical literature in \cite{MR664330,MR1001707, MR1096180} along
with projective superspace methods in the physics literature
\cite{Karlhede:1984vr,Hitchin:1986ea,deWit:2001dj,Lindstrom:2008gs}, and further tailored
for string theory applications in   \cite{Alexandrov:2008ds,Alexandrov:2008nk}.
By lifting the QK space to its twistor space, a $\CP$
bundle endowed with canonical complex and contact structures,
this method  provides an efficient parametrization of the \qk metric in terms of
a set of holomorphic transition functions between local Darboux coordinate systems, which
play a role similar to the holomorphic prepotential of special K\"ahler geometry.
Using this formalism, it was understood how to include various quantum corrections to the
metric on $\cM_H$ consistently with supersymmetry,
starting with D1-D(-1)-instantons \cite{RoblesLlana:2006is}, further including all
D-brane instantons at the fully non-linear level \cite{Alexandrov:2008gh,Alexandrov:2009zh}
and finally, NS5-brane instantons in the linear
approximation  \cite{Alexandrov:2010ca}
(see also \cite{Rocek:2005ij,Alexandrov:2006hx,RoblesLlana:2007ae,Pioline:2009qt,Bao:2009fg,Pioline:2009ia,Alexandrov:2010np,Alexandrov:2011ac}
for closely related work).
These advances took place in parallel with related developments
in the context of rigid $\cN=2$ field theories \cite{Gaiotto:2008cd,Gaiotto:2010be,Haghighat:2011xx,Neitzke:2011za}.

Unlike the calculation of the one-loop correction to $\cM_H$,
which was found by a scattering amplitude calculation
\cite{Antoniadis:1997eg,Gunther:1998sc,Antoniadis:2003sw}, the above results on instanton corrections
were obtained by postulating certain  symmetries and dualities.
Among those, the isometric action of the modular group $SL(2,\IZ)$, inherited from the
S-duality symmetry of type IIB string theory in ten dimensions played
a central role  \cite{RoblesLlana:2006is,Alexandrov:2010ca}.
Nevertheless, it has not been verified so far that the type IIA construction of
the D-instanton corrections in \cite{Alexandrov:2008gh,Alexandrov:2009zh} was
in fact consistent with S-duality of the mirror formulation. In particular, in the
limit where only D(-1), D1 and D3 instantons are retained, the metric on $\cM_H$
must be modular invariant, which provides a non-trivial constraint on the generalized
Donaldson-Thomas invariants governing such D-instantons.

One of the difficulties in demonstrating S-duality is an inherent ambiguity in the twistor
construction, namely the fact  that the Darboux coordinates are only defined up to a
local complex contact transformation. Thus, while any isometry of $\cM_H$ can always be lifted to a
holomorphic action in twistor space, the Darboux coordinates need only be covariant
up to a local complex contact transformation. Thus, the condition for S-duality
invariance is that the set of holomorphic transition functions should commute with
S-duality, up to local complex transformations in each patch.

In this work we give a solution to this problem and provide a twistorial
construction of a general class of QK metrics preserving two continuous commuting isometries,
which is explicitly $SL(2,\IZ)$-invariant and parametrized by a suitably covariant family of transition functions $G_{m,n}$.
Physically, the two isometries correspond to the unbroken Peccei-Quinn symmetries in the absence
of D5 and NS5-brane instantons. For vanishing $G_{m,n}$, these metrics reduce
to the local c-map metrics \cite{Cecotti:1988qn,Ferrara:1989ik},
which arise in the weak coupling, large volume limit of type IIB string theory.
Thus, this class of the metrics should
contain the quantum corrected HM moduli space in the limit where
fivebrane instantons  are exponentially suppressed compared
to D(-1), D1, D3 instantons. After taking a rigid limit, this class should also contain
the \hk metric on the Coulomb branch of five-dimensional $\cN=2$ gauge theories
compactified on a torus $T^2$, where the modular group of the
torus plays the role of S-duality. Thus, this construction should provide constraints on
the contributions of the monopole string instantons considered in \cite{Haghighat:2011xx}.

To motivate this construction,
we first reconsider the much simpler case of $SL(2,\IZ)$-invariant toric QK spaces,
i.e. $4n$-dimensional spaces $\cM_H$ with  $n+1$ commuting isometries. Physically, this case
arises when only  D1-D(-1)-contributions
(or equivalently $(p,q)$-string instantons) are retained and, just like the undeformed $c$-map metric,
is amenable to the Legendre transform construction. This case
was studied in detail in \cite{Alexandrov:2009qq}, where a set of {\it ad hoc}
local contact transformations was designed that bring  the general
type IIA construction of \cite{Alexandrov:2008gh}  into a manifestly
$SL(2,\IZ)$-invariant form. This invariance was then exploited in order to obtain
instanton corrections to the `mirror map',
relating the type IIB fields (covariant under S-duality) with the type IIA fields
(covariant under monodromies). However, the geometric meaning of both the contact transformation
and the quantum mirror map, had remained unclear. Here we illuminate
the {\it ad hoc} construction
of \cite{Alexandrov:2009qq}, exposing the inner workings of S-duality in the twistor space,
and considerably streamlining the derivation of the mirror map.

Having recast the construction of \cite{Alexandrov:2009qq} in a manifestly S-duality
invariant fashion, the extension to S-duality invariant QK metrics with only two commuting
isometries is then relatively straightforward, despite the fact that such metrics are
no longer amenable to the Legendre transform construction.
The key ingredient  is the non-linear
condition  \eqref{StransG} on the holomorphic transition functions $G_{m,n}$,
which ensures the  $SL(2,\IZ)$-invariance of the QK metric. Identifying
the  functions $G_{m,n}$ which are relevant in the context of
D3-brane instanton corrections to the
the HM moduli space lies beyond the scope of this paper.
In subsequent work \cite{amp-to-appear} we provide a detailed analysis of
the modular properties of the
type IIA twistorial construction of \cite{Alexandrov:2008gh},
uncovering how S-duality is realized in this framework.

The organization of the paper is as follows. In Section \ref{sec_review}
we provide a brief review of the HM moduli space both at classical and quantum level.
In particular, we explain how its QK geometry, corrected by D2-instantons of type IIA theory,
is encoded in the twistor data consisting of a covering of the twistor fiber
and an associated set of holomorphic transition functions.
Then in Section \ref{sec_d1} we study the realization of S-duality
in the dual type IIB formulation keeping contributions
only from D1-D(-1)-instantons. By refining the construction of \cite{Alexandrov:2009qq},
we demonstrate how S-duality and the quantum mirror map follow
from the simple transformation property \eqref{transGmn} of the transition functions encoding
instanton corrections.
In Section \ref{sec_thm}, we turn to the more general case of QK spaces with two commuting isometries,
and provide sufficient conditions on the covering of the twistor sphere and holomorphic
transition functions which ensure the $SL(2,\IZ)$-invariance of the metric.
We conclude with some open questions in Section \ref{sec-concl}.
Two appendices contain a review of the twistorial
description of QK spaces and a proof of the S-duality invariance of
the construction of Section \ref{sec_thm} in the linear approximation.

\section{Hypermultiplet moduli space in CY string vacua}
\label{sec_review}

In this section, we briefly recall some basic facts about the hypermultiplet moduli space
$\cM_H$ in type II string theory compactified on a CY threefold. Since mirror symmetry identifies type IIA
compactified on a Calabi-Yau threefold $\CY$ and type IIB compactified on the mirror
threefold $\CYm$, the moduli space $\cM_H$ has two equivalent, dual descriptions.
On either side, it governs the dynamics of\footnote{Our conventions
are such that the indices $\Lambda,\Sigma,\dots$ (resp. $a,b,\dots$) run from 0 (resp. 1) till $h^{2,1}(\CY)=h^{1,1}(\CYm)$.}
\begin{enumerate}
\item
the four-dimensional dilaton $r=e^\phi=1/g_4^2$ or its ten-dimensional cousin $g_s$;
\item
the complex moduli $z^a= b^a + \I t^a$ parametrizing the moduli space $\KC(\CY)$
of complex structures on $\CY$ or the complexified K\"ahler structure of  $\KK(\CYm)$ ---
these two spaces also describe the VM moduli space of the dual type II theory compactified
on the same CY threefold, and are identified by classical mirror symmetry;
\item
the RR scalars $\zeta^\Lambda, \tzeta_\Lambda$ or $c^0, c^a, \cla, \cl0$
obtained as integrals of RR gauge potentials over a symplectic basis of cycles in
$H_3(\CY,\IZ)$ or $H_{\rm even}(\CYm,\IZ)$;
\item
the NS axion $\sigma$ or $\psi$, dual to the NS 2-form $B$ in four dimensions.
\end{enumerate}
On the type IIB side, the ten-dimensional string coupling $\tau_2\equiv 1/g_s$
and the RR axion $\tau_1\equiv c^0$ combine into the ten-dimensional axio-dilaton field $\tau = \tau_1+\I \tau_2$.
As a result, the field basis $\bigl(\tau,b^a,t^a,c^a,\cla,\cl0,\psi\bigr)$ has simple transformation
rules under S-duality.
In contrast, the field basis $\bigl(\phi,z^a,\zeta^\Lambda,\tzeta_\Lambda,\sigma\bigr)$
transform naturally under monodromies, which are manifest  on the type IIA side as
they correspond to a change of symplectic basis in $H_3(\CY,\IZ)$. Thus, we refer to the
latter field basis as the type IIA coordinates, and to the former as the type IIB coordinates.

\subsection{Classical metric}

Supersymmetry constrains the moduli space $\cM_H$ to carry a \qk metric with negative
scalar curvature \cite{Bagger:1983tt}.
At classical level the metric on $\cM_H$ can be obtained from the moduli space $\KC(\CY)$
of complex structures on $\CY$, or equivalently the moduli space $\KK(\CYm)$ of
\kahler structures on $\CYm$, via the local $c$-map
construction \cite{Cecotti:1988qn,Ferrara:1989ik}. The space $\KC(\CY)=\KK(\CYm)$
is a  special \kahler manifold characterized by a
holomorphic prepotential $F(X^\Lambda)$, homogeneous of degree 2 in
the special complex coordinates $X^\Lambda$ such that $X^a/X^0=z^a=b^a+\I t^a$.
Thus, the classical metric is completely determined by the prepotential
and is given by \cite{Ferrara:1989ik}
\be
\begin{split}
\de s_{\rm cl}^2=&\frac{\de r^2}{r^2}
+4 \de s^2_{\cS\cK}
-\frac{1}{2r}\left(\de\tzeta_\Lambda - \bar\cN_{\Lambda\Sigma} \de\zeta^\Sigma\right)
\Im \cN^{\Lambda\Lambda'}\left(\de\tzeta_{\Lambda'} - \cN_{\Lambda'\Sigma'} \de\zeta^{\Sigma'}\right)
\\
&+ \frac{1}{16 r^2} \left(\de \sigma + \tzeta_\Lambda \de \zeta^\Lambda -\zeta^\Lambda \de \tzeta_\Lambda\right)^2,
\end{split}
\label{hypmettree}
\ee
where $\de s^2_{\cS\cK}$ is the metric on $\KC(\CY)=\KK(\CYm)$ with \kahler
potential $\cK=-\log[ \I ( \bar X^\Lambda F_\Lambda - X ^\Lambda \bar F_\Lambda)]$,
\be
\label{defcN}
\cN_{\Lambda\Lambda'} =\bar\tau_{\Lambda\Lambda'} +
2\I \frac{ [\Im \tau \cdot X]_\Lambda
[\Im \tau \cdot X]_{\Lambda'}}
{X^\Sigma \, \Im \tau_{\Sigma\Sigma'}X^{\Sigma'}},
\qquad
\tau_{\Lambda\Sigma}\equiv \partial_{X^\Lambda}\partial_{X^\Sigma} F(X).
\ee
Here we expressed the metric in terms of the type IIA fields where it takes an especially simple form.
In particular, it is explicitly invariant under symplectic transformations and
the continuous shifts generating the Heisenberg algebra
\be
\label{heis0}
T_{\eta^\Lambda,\tleta_\Lambda,\kappa}:\quad
\bigl(\zeta^\Lambda,\tzeta_\Lambda,\sigma\bigr)\mapsto
\bigl(\zeta^\Lambda + \eta^\Lambda ,\
\tzeta_\Lambda+ \tleta_\Lambda ,\
\sigma + 2 \kappa - \tleta_\Lambda \zeta^\Lambda
+ \eta^\Lambda \tzeta_\Lambda\bigr).
\ee
Moreover, the symplectic vector $(X^\Lambda,F_\Lambda)$ is identified as the
integral of the holomorphic three-form on $\CY$ along a symplectic basis in $H_3(\CY,\IZ)$,
while $\cN_{\Lambda\Sigma}$ and  $\tau_{\Lambda\Sigma}$ are the Weil and
Griffiths period matrices on the intermediate Jacobian $H^3(\CY,\IR)/H^3(\CY,\IZ)$,
respectively.

To rewrite the metric \eqref{hypmettree} in type IIB variables, we use the fact that
the holomorphic prepotential can be written, in the limit where the volume of
the mirror threefold $\CYm$ is infinite,  as\footnote{In general,
the cubic prepotential \eqref{Flv} must be supplemented by a quadratic
term $\frac12 A_{\Lambda\Sigma} X^\Lambda X^\Sigma$ which is necessary
for consistency with charge quantization. This term can always be removed by
working in  a non-integer symplectic basis, which we assume in this work. \label{fooquad}}
\be
\label{Flv}
F^\lvol (X^\Lambda)=-\kappa_{abc} \,\frac{X^a X^b X^c}{6 X^0},
\ee
where $\kappa_{abc}$ is the triple intersection product in $H_4(\CYm,\IZ)$.
Moreover, the type IIA fields are related to the type IIB ones via the following classical
`mirror map' \cite{Bohm:1999uk}
\be
\label{symptobd}
\begin{split}
r &=  \frac{ \tau_2^2}{2} \, \cV\, ,
\qquad
z^a=b^a+\I t^a\, ,
\qquad
\zeta^0=\tau_1\, ,
\qquad
\zeta^a = - (c^a - \tau_1 b^a)\, ,
\\
\tzeta_a &=  \cla+ \frac{1}{2}\, \kappa_{abc} \,b^b (c^c - \tau_1 b^c)\, ,
\qquad
\tzeta_0 =\, \cl0-\frac{1}{6}\, \kappa_{abc} \,b^a b^b (c^c-\tau_1 b^c)\, ,
\\
\sigma &= -2 (\psi+\frac12  \tau_1 \cl0) + \cla (c^a - \tau_1 b^a)
-\frac{1}{6}\,\kappa_{abc} \, b^a c^b (c^c - \tau_1 b^c)\, ,
\end{split}
\ee
where $\cV$ is the volume of the threefold $\CYm$ in string units,
\be
\label{defrphi}
\cV=\frac16\int_\CYm J\wedge J\wedge J
= \frac16 \kappa_{abc}t^a t^b t^c .
\ee

The virtue of the coordinate transformation \eqref{symptobd} is to make manifest
the invariance of the metric \eqref{hypmettree}  under the action of
$SL(2,\IZ)$ \cite{Gunther:1998sc,Bohm:1999uk}\footnote{In \cite{Alexandrov:2010ca}
it was found that the transformation of $\cla$ must contain an additional constant shift
proportional to the second Chern class $c_{2,a}$ of $\CYm$, which is necessary
for consistency with charge quantization. This subtlety is closely related to the
quadratic correction to the prepotential mentioned in the previous footnote, and
can be safely ignored for the purposes of this paper.}
\be\label{SL2Z}
\begin{split}
&\quad \tau \mapsto \frac{a \tau +b}{c \tau + d} \, ,
\qquad
t^a \mapsto t^a |c\tau+d| \, ,
\qquad
\cla\mapsto \cla\, ,
\\
&
\begin{pmatrix} c^a \\ b^a \end{pmatrix} \mapsto
\begin{pmatrix} a & b \\ c & d  \end{pmatrix}
\begin{pmatrix} c^a \\ b^a \end{pmatrix},
\qquad
\begin{pmatrix} \cl0 \\ \psi \end{pmatrix} \mapsto
\begin{pmatrix} d & -c \\ -b & a  \end{pmatrix}
\begin{pmatrix} \cl0 \\ \psi \end{pmatrix},
\qquad
ad-bc=1 ,
\end{split}
\ee
which corresponds to the S-duality symmetry of type IIB supergravity
in ten dimensions. The metric  \eqref{hypmettree}  is in fact invariant
under the continuous action of $SL(2,\IR)$, but as we now recall,
this invariance is broken by quantum corrections.

\subsection{Quantum corrections \label{sec_Sinst}}

Away from the classical, large volume limit, the HM moduli
 space receives two types of quantum corrections.
The first type, namely $\alpha'$-corrections, occur only in the type IIB formulation.
These effects preserve the $c$-map form \eqref{hypmettree} of the metric, and simply
correct   the holomorphic prepotential \eqref{Flv}
into $F=F^{\lvol}+F^{\ws}$ where \cite{Candelas:1990rm,Hosono:1993qy}
\be
F^{\ws}(X^\Lambda)=
\chi_\CYm\,
\frac{\zeta(3)(X^0)^2}{2(2\pi\I)^3}
-\frac{(X^0)^2}{(2\pi\I)^3}{\sum_{q_a\gamma^a\in H_2^+(\CYm)}} n_{q_a}^{(0)}\, \Li_3\left(
e^{2\pi \I  q_a X^a/X^0}\right) .
\label{Fws}
\ee
The first term, proportional to the  Euler number $\chi_\CYm$ of $\CYm$, corresponds to
a perturbative correction in the $\alpha'$-expansion around the large volume limit.
The second term corresponds to a sum of worldsheet instantons, labeled by their effective
homology class $q_a\gamma^a\in H_2^+(\CYm,\IZ)$ (i.e. $q_a\in \IZ^+$ for all $a$,
not all $q_a$'s vanishing simultaneously), and weighted by the genus zero Gopakumar-Vafa invariants
$n_{q_a}^{(0)}\in \IZ$. These effects contribute through the tri-logarithm function
$\Li_3(x)=\sum_{m=1}^\infty m^{-3}x^m$, which takes into account
multi-covering effects. Note that the last two terms
in \eqref{Fws} may be combined by including the zero class
$q_a=0$ in the sum and setting $n_0^{(0)}=-\chi_\CYm/2$.
The metric on $\cM_H$ including the $\alpha'$-corrections is expressed as in \eqref{hypmettree}
where $F$ denotes now the full prepotential.

In contrast, the second type of quantum corrections, namely
 corrections in the string coupling $g_s$, take
the metric outside the class of $c$-map metrics.
At the perturbative level, it can be argued
that the only non-trivial correction occurs at one-loop, and is determined solely
by the Euler number $\chi_\CY=-\chi_\CYm$  \cite{Antoniadis:1997eg,Gunther:1998sc,Robles-Llana:2006ez,Alexandrov:2008nk}.
Its effect on the metric was explicitly evaluated in \cite{Robles-Llana:2006ez,Alexandrov:2007ec},
and will be described in twistorial terms in Section \ref{subsec-twA}.

At the non-perturbative level, the corrections to $\cM_H$ split again in two types.
The first type corresponds, on the type IIA side,  to D2-branes wrapping special
Lagrangian cycles in $\CY$, or, on the type IIB side, to D5-D3-D1-D(-1)-branes
wrapping algebraic cycles of the corresponding dimension (or more generally,
coherent sheaves on $\CYm$). In either case, the correction takes the schematic
form  \cite{Becker:1995kb}
\be
\label{d2quali}
\delta \de s^2\vert_{\text{D2}} \sim \Omega(\gamma,z^a)\, e^{ -2\pi|Z_\gamma|/g_s
- 2\pi\I (q_\Lambda \zeta^\Lambda-p^\Lambda\tzeta_\Lambda)} ,
\ee
where the charge vector $\gamma=(p^\Lambda,q_\Lambda)=(p^0,p^a,q_a,q_0)$
takes value in $H_3(\CY,\IZ)$ or $H_{\rm even}(\CYm,\IZ)$, the central charge function
($z^0\equiv 1$)
\be
\label{defZ}
Z_\gamma(z^a) = q_\Lambda z^\Lambda- p^\Lambda F_\Lambda(z^a)
\ee
measures the area of the supersymmetric cycle in the homology class $\gamma$,
and $\Omega(\gamma,z^a)$ is the corresponding Donaldson-Thomas invariant, which counts
(with signs) the number of supersymmetric cycles with charge $\gamma$.
Due to their dependence on the RR axions $\zeta^\Lambda,\tzeta_\Lambda$,
or equivalently $\tau_1,c^a,\cla,\cl0$, the D-instanton effects \eqref{d2quali}
break the continuous Heisenberg symmetries \eqref{heis0} to a discrete
subgroup. In the large volume limit on the type IIB side, this breaking occurs
in a hierarchical fashion, namely translations along $\tau_1$ are broken
at leading order, translations along $c^a$ at order $\cO(e^{-\cV^{1/3}})$, translations
along $\cla$ at order $\cO(e^{-\cV^{2/3}})$ and finally  translations
along $\cl0$ at order $\cO(e^{-\cV})$.

Finally, the last type of corrections is generated by Euclidean NS5-branes
wrapping the whole CY threefold. They contribute to the metric schematically  as \cite{Becker:1995kb}
\be
\delta \de s^2\vert_{\text{NS5}}  \sim
e^{-2 \pi |k| \cV /g_s^2-\I\pi  k \sigma},
\label{couplNS5}
\ee
and break the last continuous isometry along $\sigma$ (or $\psi$) which remained
after inclusion of all D-instanton effects. Just like D5-branes, they are suppressed
as $\cO(e^{-\cV})$ in the large volume limit.

What is the fate of S-duality at the quantum level? Firstly, the
continuous $SL(2,\IR)$ symmetry \eqref{SL2Z}
is certainly broken by the $\alpha'$-corrections in \eqref{Fws}.
However, this symmetry descends from the usual S-duality action in ten-dimensional
string theory, and there is overwhelming evidence that the modular subgroup
$SL(2,\IZ)$ remains a symmetry at the quantum level (see \cite{Green:1997tv,Bohm:1999uk} and
citations thereof). It is a priori unclear whether this  symmetry  should remain intact
after compactification on a Calabi-Yau three-fold $\CYm$, but this is supported by the fact
that upon compactifying on $S^1$, T-dualizing to type IIA string theory and lifting to
M-theory, this action becomes the usual large diffeomorphisms of the torus in M-theory
on $\CYm\times T^2$. Thus, it is natural to assume that quantum corrections should preserve
an isometric action of the discrete group $SL(2,\IZ)$.

Furthermore, from \eqref{SL2Z} it is possible to distinguish two sectors
which should be preserved by S-duality. The first one is obtained by setting all fivebrane
charges to zero, i.e. ignoring effects of order $\cO(e^{-\cV})$ in the large volume limit,
leaving only D3-D1-F1-D(-1)-instantons.
As noted above, in this approximation the moduli space has two continuous isometries
along $\psi$ and $\cl0$.
The second sector arises by further setting the D3-brane charge $p^a=0$, i.e.
ignoring effects of order $\cO(e^{-\cV^{2/3}})$ or smaller in the large volume limit,
leaving only D1-F1-D(-1)-instantons. In this limit the number of isometries is increased to
$h^{1,1}(\CYm)+2=n+1$, where $4n$ is the real dimension of $\cM_H$.
The aim of this paper is to provide a manifestly S-duality invariant description of these two sectors.

\subsection{Twistorial description of $\cM_H$\label{subsec-twA}}

To describe instanton corrections to the classical HM geometry,
consistently with supersymmetry, it is indispensable
to use the twistor description of QK spaces, very briefly summarized in Appendix \ref{secQK}.
However, before going to the non-perturbative physics, one should understand how
the perturbative moduli space is encoded in this formalism. This has been understood
in \cite{Alexandrov:2008nk,Alexandrov:2008gh}, based on the previous results of \cite{Rocek:2005ij,Neitzke:2007ke}.

As explained in Appendix \ref{secQK}, the twistor approach allows to
encode a QK metric in terms of
a set of holomorphic functions $\Hij{ij}$. They describe contact transformations between
local Darboux coordinate systems for the complex contact one-form \eqref{contform},
which are attached to various patches $\cU_i$ of an open covering of the twistor fiber $\CP$.
To describe the perturbative metric on $\cM_H$, it is sufficient to cover $\CP$
by two patches $\cU_+$, $\cU_-$ centered around the north and south poles,
$\varpi=0$ and $\varpi=\infty$, and a third patch\footnote{The patch $\cU_0$
could in principle be omitted but is very convenient for exposing symplectic invariance.}
 $\cU_0$ which surrounds the equator.
The transition functions between these  patches are given in
terms of the holomorphic prepotential $F(X^\Lambda)$ by
\be
\label{symp-cmap}
\hHij{+0}= F(\xi^\Lambda) ,
\qquad
\hHij{-0}=\bF(\xi^\Lambda) .
\ee
Using \eqref{txiqline}, we find that the Darboux coordinates in the patch $\cU_0$
are given by \cite{Neitzke:2007ke,Alexandrov:2008nk}
\be
\label{gentwi}
\begin{array}{rcl}
\xi^\Lambda &=& \zeta^\Lambda + \frac12\,\tau_2
\left( \varpi^{-1} z^{\Lambda} -\varpi \,\bz^{\Lambda}  \right) ,
\\
\txi_\Lambda &=& \tzeta_\Lambda + \frac12\,\tau_2
\left( \varpi^{-1} F_\Lambda-\varpi \,\bF_\Lambda \right) ,
\\
\tilde\alpha&=& \sigma + \frac12\,\tau_2
\left(\varpi^{-1} W -\varpi \,\bar W \right)
+\frac{\I\chi_\CY}{24\pi} \,\log \varpi \, ,
\end{array}
\ee
where $W$ denotes the `superpotential'
\be
\label{defW}
W(z) =  F_\Lambda(z) \, \zeta^\Lambda - z^\Lambda \tzeta_\Lambda .
\ee
The last, logarithmic term in $\tilde\alpha$ encodes the effect of the one-loop $g_s$-correction
and corresponds to an anomalous dimension $c=\chi_\CY / (96\pi)$ in the language
of Appendix A.
In \eqref{gentwi}, we have traded the Darboux coordinate $\alpha$ from Appendix \ref{secQK}
for $\tilde \alpha \equiv -2\alpha - \txi_\Lambda \xi^\Lambda$, and the integration constants
$A^\Lambda,B_\Lambda, B_\alpha$ for
\be
\zeta^\Lambda= A^\Lambda,
\qquad
\tzeta_\Lambda=
B_\Lambda + \Re F_{\Lambda\Sigma} A^\Sigma ,
\qquad
\sigma = -2 B_\alpha - A^\Lambda B_\Lambda .
\label{relTAf}
\ee
Moreover, the contact potential $\Phi$ computed from \eqref{solcontpot}
is related to the four- and ten-dimensional string couplings by
\be
r=e^{\Phi}=\frac{\tau_2^2}{16}\, e^{-\cK}+\frac{\chi_{\CY}}{192\pi},
\ee
which generalizes the first relation in \eqref{symptobd}.

The way to incorporate D-instanton corrections to the above twistor formulation of $\cM_H$
was explained in \cite{Alexandrov:2008gh,Alexandrov:2009zh}, in close analogy
with the field theory construction of \cite{Gaiotto:2008cd}.
We shall refer to this construction as the `type IIA construction', as it is manifestly invariant under symplectic
transformations (i.e. monodromies), which are manifest on the type IIA side.
Later in this paper, we shall encounter a different `type IIB construction' of the same twistor space,
which makes S-duality manifest at the expense of obscuring symplectic transformations.

To explain this construction, one should first recall that, instead of covering the twistor
space by open patches surrounded by closed contours, it is possible to consider a set
of open contours with associated holomorphic transition functions across them (see Appendix \ref{secQK}).
For a fixed value of the moduli $z^a$ and
any state of charge $\gamma$ with $\Omega(\gamma,z^a)\neq 0$, we then consider the BPS ray
\be
\ell_\gamma = \{ \varpi\in \CP\ :\ \ Z_\gamma(z^a)/\varpi\in \I \IR^-\},
\label{BPSray}
\ee
where $Z_\gamma(z^a)$ is the same central charge function as in \eqref{defZ}.
These rays extend from the north to the south pole and divide the patch $\cU_0$ into
angular sectors. Across each BPS ray $\ell_\gamma$, the Darboux coordinates
are required to jump by the complex contact transformation generated
by  \cite{Alexandrov:2009zh}
\be
H_\gamma=G_{\gamma}-\hf\,q_{\Lambda} p^\Lambda (G_{\gamma}')^2 ,
\qquad
G_\gamma(\Xi_\gamma)=\frac{\hnkl}{(2\pi)^2}\,
\Li_2\left(\sigma_{\gamma} \, e^{-2\pi \I \Xikl} \right),
\label{transellg}
\ee
where $G'_\gamma$ denotes the derivative with respect to $\Xi_\gamma= q_\Lambda \xii{\gamma}^\Lambda-p^\Lambda \txii{\gamma}_\Lambda$,
$\Omega(\gamma)$ is the Donaldson-Thomas invariant\footnote{The fact that these invariants also depend on the moduli $z^a$
does not spoil the holomorphicity of \eqref{transellg} because they are piecewise constant.
Across a wall of marginal stability, the $\Omega$'s jump and the
BPS rays exchange their order in such a way that the metric on $\cM_H$ remains continuous \cite{Gaiotto:2008cd,Alexandrov:2011ac}.},
and $\sigma_\gamma$ is a certain phase factor depending only on the charge, known as a quadratic refinement.
The perhaps unfamiliar formula \eqref{transellg} is designed so as to generate
jumps of the Darboux coordinates $(\xi^\Lambda,\txi_\Lambda)$ across $\ell_\gamma$ given by the usual symplectomorphism \cite{ks}
\be
\begin{split}
\Delta \xi^\Lambda =&\, \frac{1}{2\pi\I}\, \Omega(\gamma)\, p^\Lambda\, \log\( 1-\sigma_\gamma \, e^{-2\pi \I \Xi_{\gamma}}\) ,
\\
\Delta \txi_\Lambda =&\, \frac{1}{2\pi\I}\, \Omega(\gamma)\, q_\Lambda\, \log\( 1-\sigma_\gamma \, e^{-2\pi \I \Xi_{\gamma}}\).
\end{split}
\label{elemks}
\ee
The additional term proportional to $(G_{\gamma}')^2$ in \eqref{transellg} arises upon
integrating this symplectomorphism to a generating function  $\Hij{ij}$,
which depends on the coordinate $\xii{i}^\Lambda$ in one patch and the dual coordinate
$\txii{j}_\Lambda$ in the other patch (see Appendix \ref{secQK}).

Unfortunately, except when all D-instantons are mutually local (i.e. $\langle \gamma,\gamma'\rangle=0$ for any pair
with non-vanishing $\Omega(\gamma),\Omega(\gamma')$),
the Darboux coordinates determined by these
gluing conditions cannot be expressed in closed form as in \eqref{gentwi}.
Instead, they are determined
by integral equations \eqref{txiqline} which upon substitution of \eqref{transellg} take the form
of a Thermodynamic Bethe Ansatz \cite{Gaiotto:2008cd,Alexandrov:2010pp}.
Moreover, while the construction outlined above is manifestly covariant under
symplectic transformations, it is not
manifestly invariant under S-duality. In the presence of D5-brane instantons (i.e. $p^0\neq 0$),
this is of course expected since S-duality mixes D5-branes with NS5-branes, which
are not included in the construction above. In the absence of D5-branes however
(more precisely, in the large volume limit where both D5-branes and NS5-branes
can be ignored), the metric should be invariant under the discrete group $SL(2,\IZ)$,
yet this is far from obvious  from the construction above.
This invariance has been shown in the D1-D(-1) sector \cite{Alexandrov:2009qq},
extending the results of  the earlier work \cite{RoblesLlana:2006is}, and our goal
is to extend this construction to D3-D1-D(-1)-instantons. Before doing
so however, we return to the simpler case of D1-D(-1)-instantons, improving on
the earlier construction in  \cite{Alexandrov:2009qq}.

\section{S-duality and D1-D(-1)-instantons \label{sec_d1}}

In this section, we revisit the construction of the D1-D(-1)-instanton corrected metric
on the HM moduli space in type IIB Calabi-Yau vacua, emphasizing how S-duality is realized in twistor space.
In the process we considerably streamline the derivation of the `quantum mirror map' obtained in
\cite{Alexandrov:2009qq}.

\subsection{S-duality in the classical twistor space \label{subsec_classS}}

Let us first recall how S-duality is realized in the twistor space $\cZ$ of $\cM_H$ in the classical,
infinite volume limit.
This twistor space is described by the Darboux coordinates \eqref{gentwi},
upon  dropping the last logarithmic term in $\tilde\alpha$ and restricting the prepotential
to its large volume limit \eqref{Flv}.
To express them in terms of the type IIB fields,
it is sufficient to
substitute the classical mirror map \eqref{symptobd}.

An important feature of the twistor space construction is that
all isometries of a QK manifold can be lifted to a holomorphic action on $\cZ$.
In particular, the  $SL(2,\IR)$  symmetry \eqref{SL2Z} ought to
be realized as a holomorphic action  on the Darboux coordinates
$\xi^\Lambda,\txi_\Lambda$ and $\alpha$. This indeed turns out to be the case provided
the fiber coordinate $\varpi$ transforms as
\be
\varpi  \mapsto  \frac{1+\tzm\varpi}{\tzm-\varpi}=
-\frac{\tzp-\varpi}{1+\tzp\varpi} ,
\label{SL2varpi}
\ee
where $\tzpm$ denote the two roots of the quadratic equation $c\xi^0(\varpi)+d=0$,
\be
\tzpm = \frac{ c \tau_1 + d \mp | c\tau + d |}{c \tau_2} ,
\qquad
\tzp \tzm = -1.
\label{poles}
\ee
Combining \eqref{SL2varpi} with \eqref{SL2Z}
and the expressions \eqref{gentwi} for the Darboux coordinates  in terms of the type IIB fields
then leads to the following non-linear holomorphic action on the Darboux coordinates
in the patch $\cU_0$ \cite{Alexandrov:2008gh}\footnote{In general, since S-duality acts on the twistor fiber, it will map
Darboux coordinates in one patch to Darboux coordinates in another patch. The virtue of the
patch $\cU_0$ is that it is mapped to itself under S-duality since it contains the points $t=\pm\I$,
see the remark at the end of this subsection.}
\be
\label{SL2Zxi}
\begin{split}
&
\xi^0 \mapsto \frac{a \xi^0 +b}{c \xi^0 + d} , \qquad
\xi^a \mapsto \frac{\xi^a}{c\xi^0+d} , \qquad
\txi_a \mapsto \txi_a +  \frac{ c}{2(c \xi^0+d)} \kappa_{abc} \xi^b \xi^c-c_{2,a}\eps(\gsl),
\\
&
\begin{pmatrix} \txi_0 \\ \alpha \end{pmatrix} \mapsto
\begin{pmatrix} d & -c \\ -b & a  \end{pmatrix}
\begin{pmatrix} \txi_0 \\  \alpha \end{pmatrix}
+ \frac{1}{6}\, \kappa_{abc} \xi^a\xi^b\xi^c
\begin{pmatrix}
c^2/(c \xi^0+d)\\
-[ c^2 (a\xi^0 + b)+2 c] / (c \xi^0+d)^2
\end{pmatrix} .
\end{split}
\ee
Under this  action,  the complex contact one-form \eqref{contform} transforms by an overall holomorphic factor
$\hCX\ui{0}\mapsto \hCX\ui{0}/(c\xi^0+d)$, leaving the complex
contact structure invariant. Furthermore, the contact potential
$e^\Phi$ transforms with modular weight $(-\frac12,-\frac12)$,
which ensures that the \kahler potential \eqref{Knuflat} varies by a \kahler
transformation,
\be
\label{SL2phi}
e^\Phi \mapsto \frac{e^\Phi}{|c\tau+d|}\, ,
\qquad
K_\cZ\mapsto  K_\cZ - \log(|c\xi^0+d| ).
\ee
These properties ensure that S-duality is indeed a symmetry of the classical twistor space.

To illuminate the action \eqref{SL2varpi} of S-duality on the fiber, it is useful to make a
Cayley transformation of the fiber coordinate and define\footnote{We are grateful to J. Manschot
for suggesting this redefinition.}
\be
\label{Cayley}
z=\frac{t+\I}{t-\I}\, ,
\qquad
t=-\I\, \frac{1+z}{1-z}\, .
\ee
Under \eqref{SL2varpi}, this new coordinate transforms by
the compensating $U(1)$ rotation induced by
the right action on the coset $U(1)\backslash SL(2,\IR)$,
\be
\label{ztrans}
z\mapsto \frac{c\bar\tau+d}{|c\tau+d|}\, z ,
\ee
i.e. $z$ has modular weight $(-\tfrac12,\tfrac12)$.
In particular,  the points $z=0,\infty$ (corresponding to $t=\pm\I$)
stay invariant under $SL(2,\R)$, whereas the zeros \eqref{poles}
of $c\xi^0+d$ are now given by
$ z^{c,d}_\pm =\mp \sqrt{\frac{c\tau+d}{c\bar\tau+d}} $. Rewriting
 the Darboux coordinate $\xi^0$ in terms of $z$,
\be
\xi^0 = \tau_1+ \I \tau_2 \, \frac{z^{-1}+z}{z^{-1}-z}\, ,
\ee
we recognize in $\xi^0$ and $z$ the analog of the  'canonical' and 'twisting' parameters
 introduced in \cite{Kapustin:2006pk}.

\subsection{Type IIB twistorial construction of D1-D(-1)-instanton corrections \label{sec_twid1}}

As discussed in Section \ref{sec_Sinst}, upon including worldsheet instanton corrections
and the one-loop $g_s$-correction to the metric \eqref{hypmettree}, the continuous $SL(2,\IR)$
action \eqref{SL2Z} is no longer isometric. Nevertheless, it was shown in \cite{RoblesLlana:2006is}
that  invariance under a discrete subgroup $SL(2,\IZ)$ may be restored
by incorporating D1 and D(-1)-instanton corrections.
This was achieved in the framework of projective superspace
by constructing a modular invariant completion of the \hk potential,
a close cousin of our contact potential $e^\Phi$ defined on the Swann bundle \cite{MR1096180} over $\cM_H$.
At the same time,
these instanton corrections should agree with the type IIA construction presented
in Section \ref{subsec-twA} restricted to the D1-D(-1) sector, with $p^\Lambda=0$.
This equivalence was demonstrated in \cite{Alexandrov:2009qq},
where the construction of  \cite{RoblesLlana:2006is} was translated in
the twistorial language, and shown to be related to the type IIA construction by a certain
complex contact transformation. Our aim is to revisit the twistorial construction of
\cite{Alexandrov:2009qq} and expose its invariance under S-duality.

As explained in \cite{Alexandrov:2009qq}, the projective superspace construction
of  \cite{RoblesLlana:2006is} can be cast in the twistorial language by using
a covering of the $\CP$ fiber by six patches
\be
\label{patch-IIB}
\cZ=\cU_+\cup\cU_-\cup\cU_{0^+}\cup\cU_{0^-}  \cup\cU_{\IR^+}\cup\cU_{\IR^-}.
\ee
Here $\cU_\pm$ are, as usual, the patches around the north and south poles,
$\cU_{\IR^\pm}$ surround the positive and negative real half-axes,
and the patches $\cU_{0^\pm}$ cover the upper and lower half-planes, in particular,
containing the S-duality invariant points $\varpi=\pm\I$.
The transition functions between these patches
are taken to be
\be
\label{trans-funIIB}
\Hij{+0^{\pm}}=F(\xi^\Lambda) ,
\qquad
\Hij{-0^{\pm}}=\bF(\xi^\Lambda) ,
\qquad
\Hij{\IR^+ 0^{\pm}}= G_{\rm IIB}(\xi^\Lambda),
\qquad
\Hij{\IR^- 0^{\pm}}= \bG_{\rm IIB}(\xi^\Lambda),
\ee
with
\be
G_{\rm IIB}(\xi^\Lambda)
= -\frac{\I}{(2\pi)^3}\!\!\sum_{q_a\gamma^a\in H_2^+(\CYm)\cup\{0\}}\!\!\!\! n_{q_a}^{(0)}\,
\sum_{n\in \IZ \atop m>0} \frac{e^{-2\pi \I m q_a\xi^a}}{m^2(m\xi^0+n)},
\label{defGIIB}
\ee
where $n_{q_a}^{(0)}$ are the same genus zero Gopakumar-Vafa invariants
which govern the worldsheet instantons \eqref{Fws}.

It is important to note that the transition functions $G_{\rm IIB}$
introducing instanton corrections are independent of $\txi_\Lambda$ and $\alpha$.
This is necessary for the existence of the $n+1$ commuting isometries corresponding
to translations along $\cla,\cl0$ and $\psi$, or equivalently holomorphic translations
along $\txi_\Lambda$ and $\alpha$. By the moment map construction \cite{MR872143},
these $n+1$ isometries lead to the existence of $n+1$ $\cO(2)$-valued sections modulo
rescalings, which correspond to the Darboux coordinates $(1,\xi^\Lambda)$. Thus
the coordinates $\xi^\Lambda$ are globally well-defined (up to rescalings) and given
by their perturbative expressions \eqref{gentwi} in all patches.

The remaining Darboux coordinates, $\txi_\Lambda$ and $\alpha$, determined by the transition functions \eqref{trans-funIIB}
were computed explicitly in \cite{Alexandrov:2009qq}.
In particular, it was shown that it is possible to relate the
coordinates $A^\Lambda,B_\Lambda,B_\alpha$
appearing in \eqref{txiqline} and the type IIB fields in such a way
that the Darboux coordinates in the patch $\cU_{0^\pm}$
transform under the combined action of \eqref{SL2Z} and \eqref{SL2varpi}
according to the {\it classical} laws \eqref{SL2Zxi}.
This ensures that the twistor space $\cZ$ and the original
moduli space $\cM_H$ carry an isometric action of $SL(2,\IZ)$. Moreover,
it was also shown that this `type IIB' twistorial construction is related to the usual
`type IIA' construction presented in Section \ref{subsec-twA} by a set
of complex contact transformations. We further discuss and clarify this relation
in Subsection  \ref{sec_equiv}.

While these results establish the invariance of the D1-D(-1)-instanton corrected
metric under S-duality,  it is fair to say that the computation of the quantum corrected
mirror map in  \cite{Alexandrov:2009qq} was rather indirect and unilluminating.
Moreover, it is highly desirable to understand how S-duality is realized
at the level of transition functions, without having to compute the Darboux
coordinates first and then check S-duality invariance. A more conceptual
understanding of these two problems
is certainly required in order to address D3-brane corrections, since unlike the
D1-D(-1) case, the integral equations \eqref{txiqline} cannot be solved in closed
form as soon as D3-instantons are included.
In the rest of this section, we disassemble the previous construction
into its bare parts and expose the inner workings of S-duality in twistor space.

\subsection{The key observations}

To understand why the above construction produces a modular invariant twistor space,
it is useful to separate the classical and instantonic parts of
the construction, and examine how S-duality reshuffles the latter.
To this end, let us refine the covering \eqref{patch-IIB}
and consider (see Fig. \ref{fig-D1inst})
\be
\label{Zpatch}
\cZ=\cU_+\cup\cU_-\cup\cU_{0^+}\cup\cU_{0^-}  \cup\(\cup_{m,n}' \cU_{m,n}\),
\ee
where $m,n$ run over all pairs of integers\footnote{This set of patches is highly
redundant, since $\cU_{km,kn}=\cU_{m,n}$ for any positive integer $k$. One could
avoid this redundancy by restricting to coprime integers $(m,n)$, at the cost of
introducing tri-logarithms in the transition functions below, but we prefer to keep
it for the sake of simplicity.
In particular, note that $\cU_{0,k}=\cU_+$ and $\cU_{0,-k}=\cU_-$.}
different from $(0,0)$ and
$\cU_{m,n}$ is an open set around the point $t_+^{m,n}$ \eqref{poles},
which is one of the zeros of $m\xi^0+n$.
The  transition functions between these patches are chosen as
\be
\label{transfun}
\Hij{+0^{\pm}}=\Fcl(\xi^\Lambda) ,
\qquad
\Hij{-0^{\pm}}=\bFcl(\xi^\Lambda) ,
\qquad
\Hij{(m,n)0^{\pm}}= G_{m,n}^{\rm D1}(\xi^\Lambda),
\ee
where $\Fcl(X)$ is the tree-level, large volume prepotential \eqref{Flv} and
\be
G_{m,n}^{\rm D1}(\xi^\Lambda)
= -\frac{\I}{(2\pi)^3}\!\!\sum_{q_a\gamma^a\in H_2^+(\CYm)\cup\{0\}}\!\!\!\! n_{q_a}^{(0)}\,
\begin{cases}
\displaystyle\frac{e^{-2\pi \I m q_a\xi^a}}{m^2(m\xi^0+n)}, &
\quad  m\ne 0,
\\
\displaystyle (\xi^0)^2 \,\frac{e^{2\pi \I n q_a\xi^a/\xi^0}}{n^3}, &\quad  m=0.
\end{cases}
\label{defG1}
\ee
Thus, we simply introduced a distinct patch for each term in the sum over $(m,n)$ in \eqref{defGIIB},
centered around the point where this term is singular. In addition, we have split
the prepotential $F$ into its classical part $\Fcl$ and its quantum part
$F^{\ws}=\sum_{n>0}G_{0,n}^{\rm D1}$, which arises in this construction as the contribution from
$m=0$ in \eqref{defG1}.

\lfig{The covering and transition functions incorporating D1-D(-1)-instantons.}{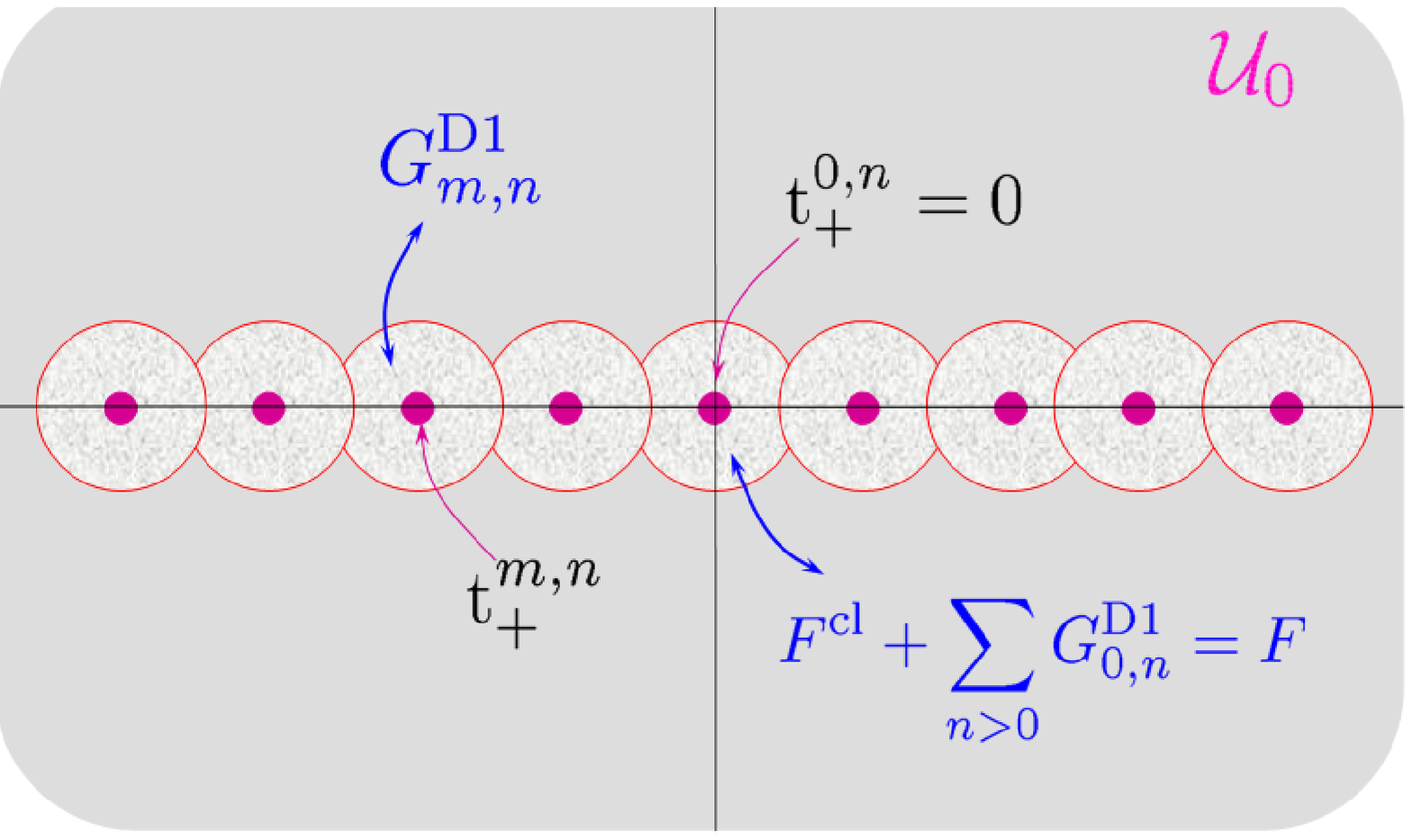}{12cm}{fig-D1inst}{-0.6cm}

The first two transition functions in \eqref{transfun} are the familiar ones arising
in the twistorial construction of the tree-level metric, which is
known to be S-duality invariant (see Section \ref{subsec_classS}).
Therefore it is sufficient to concentrate
on the patches $\cU_{m,n}$ and transition functions $G_{m,n}^{\rm D1}$.
The first crucial observation is that the patches $\cU_{m,n}$ are mapped into each other
under $SL(2,\IZ)$-transformations \eqref{SL2varpi} according to
\be
\label{SL2ZU}
\cU_{m,n}\mapsto \cU_{m',n'},
\qquad
\( m'\atop n'\) =
\(
\begin{array}{cc}
a & c
\\
b & d
\end{array}
\)
\( m \atop n \),
\ee
whereas the patches $\cU_{0^\pm}$ stay invariant.

It is natural to expect that the functions $G_{m,n}^{\rm D1}$ should similarly be
reshuffled under the transformations generated by \eqref{SL2Zxi}.
This turns out to be true, up to an important subtlety:  the transition functions $G_{m,n}^{\rm D1}$
transform into each other under $SL(2,\IZ)$, up to an overall factor of $(c\xi^0+d)^{-1}$ and
{\it up to terms which are regular in the patch $\cU_{m',n'} $},\footnote{For example,
if both $m$ and $m'=dm-cn$ are non zero,
$$
\frac{e^{-2\pi \I m q_a\xi^a}}{m^2(m\xi^0+n)}\mapsto
\frac{e^{-2\pi \I  q_a\xi^a\(m'-c\,\frac{m'\xi^0+n'}{c\xi^0+d}\)}}{(c\xi^0+d)(m'\xi^0+n')\(m'-c\,\frac{m'\xi^0+n'}{c\xi^0+d}\)^2}
=\frac{e^{-2\pi \I m' q_a\xi^a}}{m'^2(c\xi^0+d)(m'\xi^0+n')}+O\((m'\xi^0+n')^0\).
$$}
\be
G_{m,n}^{\rm D1}\mapsto\frac{G_{m',n'}^{\rm D1}}{c\xi^0+d}+\mbox{regular at } \varpi_+^{m',n'}.
\label{transGmn}
\ee
This property is sufficient for S-duality because the regular terms appearing on the r.h.s.
of \eqref{transGmn} can be canceled by a local contact transformation and do not
contribute to the contour integrals in Eqs. \eqref{txiqline}.

Yet, a further difficulty is the
fact  the integration kernel $\frac{\varpi' + \varpi}{\varpi' - \varpi}\,\frac{\de \varpi'}{\varpi'}$
appearing in these equations  transforms in a complicated fashion, which
seems to jeopardize S-duality.
Our second key observation is that
this kernel can be supplemented with a $\varpi$-independent term
\be
\label{modker}
\frac12 \,\frac{\varpi' + \varpi}{\varpi' - \varpi}\,\frac{\de \varpi'}{\varpi'}
\quad\to\quad
K(\varpi,\varpi') \frac{\de t'}{t'}\equiv
\frac12\(\frac{\varpi'+\varpi}{\varpi'-\varpi}+\frac{1/\varpi'-\varpi'}{1/\varpi'+\varpi'}\)
\frac{\de \varpi'}{\varpi'}\, ,
\ee
in such a way that the resulting kernel becomes $SL(2,\IZ)$-invariant!
Indeed, this invariance becomes obvious after expressing \eqref{modker} in terms of
the Cayley-rotated coordinate $z$ \eqref{Cayley},
\be
K(\varpi,\varpi') \frac{\de t'}{t'}
=\frac{(1+\varpi\varpi')}{(\varpi'-\varpi)(1/\varpi'+\varpi')}\frac{\de \varpi'}{\varpi'}
=\frac12\,\frac{z'+z}{z'-z}\, \frac{\de z'}{z'}\, .
\ee
Since the difference between the two kernels is $\varpi$-independent,
it can be absorbed into a redefinition of the constant terms in the $\varpi$-expansion
of Darboux coordinates. Thus, to write them in terms of the new kernel,
it is sufficient to redefine $\zeta^\Lambda,\tzeta_\Lambda$ and $\sigma$ and, as we will see in the next subsection,
this redefinition is the origin of quantum corrections to the mirror map.

\subsection{The quantum mirror map revisited}

Now we are ready to demonstrate the explicit invariance under $SL(2,\IZ)$ without
evaluating explicitly the Penrose type integrals determining the instanton contributions
and to derive a simple expression for the mirror map.
To this end, we write the integral expressions \eqref{txiqline} for Darboux coordinates by separating
the classical contributions generated by $\Hij{\pm0^\pm}$ from the quantum corrections coming from $\Hij{(m,n)0^\pm}$.
Since $\xi^\Lambda$ are uncorrected, it is sufficient to consider\footnote{Here and henceforth
the primes on sums over $m,n$ will denote that the value $m=n=0$ is omitted.}
\be
\begin{split}
\txii{0^\pm}_\Lambda& \,= \txi^{\rm cl}_\Lambda+\Delta\tzeta_\Lambda -
\half\,  {\sum_{m,n}}' \oint_{C_{m,n}} \frac{\de \varpi'}{2 \pi \I \varpi'} \,
\frac{\varpi' + \varpi}{\varpi' - \varpi}
\, \p_{\xi^\Lambda}G_{m,n}^{\rm D1}(\varpi'),
\\
\ai{0^\pm}& \,= \alpha^{\rm cl}-\frac12 \Delta\sigma
- \half \, {\sum_{m,n}}' \oint_{C_{m,n}} \frac{\de \varpi'}{2 \pi \I \varpi'} \,
\frac{\varpi' + \varpi}{\varpi' - \varpi}
\(1-\xi^\Lambda\p_{\xi^\Lambda}\)G_{m,n}^{\rm D1}(\varpi') ,
\end{split}
\label{txiqlineIIB}
\ee
where we denoted by $\txi_\Lambda^\lvol$, $\alpha^\lvol$
the Darboux coordinates \eqref{gentwi} with the prepotential replaced by its large volume limit $F^\lvol(X^\Lambda)$
and the logarithmic one-loop correction dropped.
Nevertheless, the type IIA fields $\tzeta_\Lambda$ and $\sigma$ entering  $\txi_\Lambda^\lvol$ and $\alpha^\lvol$
are still defined by relations \eqref{relTAf} with the {full}
prepotential $F$.\footnote{The relations \eqref{relTAf} are not affected by D-instantons with vanishing  $p^\Lambda$
\cite{Alexandrov:2008gh} so that they provide the correct definitions of the type IIA fields in our case. For $p^\Lambda\neq 0$,
they receive additional instanton contributions, see Eq. (3.10) in \cite{Alexandrov:2009zh}.}
This fact is at the origin of two additional contributions in \eqref{txiqlineIIB},  given by
\be
\begin{split}
\Delta \tzeta_\Lambda &\,=
- \zeta^\Sigma \Re{\sum_{n>0}} \p_{z^\Lambda}\p_{z^\Sigma}G_{0,n}^{\rm D1}(z^\Lambda),
\\
\Delta\sigma &\,=
- \zeta^\Lambda\zeta^\Sigma \Re{\sum_{n>0}} \p_{z^\Lambda}\p_{z^\Sigma}G_{0,n}^{\rm D1}(z^\Lambda).
\end{split}
\ee

Next, we perform the kernel replacement \eqref{modker} in \eqref{txiqlineIIB}, thus rewriting
the Darboux coordinates in the following form
\be
\begin{split}
\txii{0^\pm}_\Lambda&\, = \left[ \txi^{\rm cl}_\Lambda - \tzeta^{\rm inst}_\Lambda \right]-
 {\sum_{m,n}}' \oint_{C_{m,n}} \frac{\de \varpi'}{2 \pi \I \varpi'} \,K(\varpi,\varpi')
\, \p_{\xi^\Lambda}G_{m,n}^{\rm D1}(\varpi'),
\\
\ai{0^\pm}&\, = \left[\alpha^{\rm cl} +\frac12 ( \sigma^{\rm inst} + \tzeta^{\rm inst}_\Lambda \zeta^\Lambda) \right]-
{\sum_{m,n}}' \oint_{C_{m,n}} \frac{\de \varpi'}{2 \pi \I \varpi'} \,K(\varpi,\varpi')
\(1-\xi^\Lambda\p_{\xi^\Lambda}\)G_{m,n}^{\rm D1}(\varpi'),
\end{split}
\label{twlineIIBb}
\ee
where we combined all $\varpi$-independent  contributions into
\be
\begin{split}
\tzeta^{\rm inst}_\Lambda & = -
\half\,  {\sum_{m,n}}' \oint_{C_{m,n}} \frac{\de \varpi'}{2 \pi \I \varpi'} \,
\frac{1/\varpi'-\varpi'}{1/\varpi'+\varpi'}
\, \p_{\xi^\Lambda}G_{m,n}^{\rm D1}(\varpi')-\Delta \tzeta_\Lambda ,
\label{qmirmapb}
\\
\sigma^{\rm inst} & =-\zeta^\Lambda\tzeta^{\rm inst}_\Lambda
+ {\sum_{m,n}}' \oint_{C_{m,n}} \frac{\de \varpi'}{2 \pi \I \varpi'} \,
\frac{1/\varpi'-\varpi'}{1/\varpi+\varpi}\,
\(1-\xi^\Lambda\p_{\xi^\Lambda}\)G_{m,n}^{\rm D1}(\varpi')-\Delta\sigma.
\end{split}
\ee

Finally, using the property \eqref{transGmn}, it is easy to show that the
derivatives of $G_{m,n}^{\rm D1}$,
 appearing in \eqref{twlineIIBb} transform under S-duality as
\be
\begin{split}
\p_{\xi^a}G_{m,n}^{\rm D1} &\ \mapsto\ \p_{\xi^a}G_{m',n'}^{\rm D1}+\mbox{reg.}
\\
\p_{\xi^0}G_{m,n}^{\rm D1} &\ \mapsto\ d\, \p_{\xi^0}G_{m',n'}^{\rm D1}-c\, (1-\xi^\Lambda\p_{\xi^\Lambda})G_{m',n'}^{\rm D1}
+\mbox{reg.}
\\
(1-\xi^\Lambda\p_{\xi^\Lambda})G_{m,n}^{\rm D1} & \ \mapsto\ a\, (1-\xi^\Lambda\p_{\xi^\Lambda})G_{m',n'}^{\rm D1}
-b\, \p_{\xi^0}G_{m',n'}^{\rm D1}+\mbox{reg.}
\end{split}
\label{transderGmn}
\ee
Using this result and noting that the regular contributions disappear under the contour integrals,
the second term in the expression \eqref{twlineIIBb} for
$\txi_a$ is manifestly invariant, whereas the second terms in $(\txi_0,\alpha)$
transform as a doublet under S-duality,
consistently with the classical transformation properties \eqref{SL2Zxi}.
Thus, to establish that the twistor space carries a holomorphic action of $SL(2,\IZ)$,
it remains to ensure that the first terms in  \eqref{twlineIIBb}
transform as the Darboux coordinates in the classical limit.
For this purpose, it suffices to modify the
classical mirror map relations \eqref{symptobd} into
\be
\label{symptobdc}
\begin{split}
\zeta^0&=\tau_1\, ,
\qquad
\zeta^a = - (c^a - \tau_1 b^a)\, ,
\\
\tzeta_a &=  \cla+ \frac{1}{2}\, \kappa_{abc} \,b^b (c^c - \tau_1 b^c)+\tzeta_a^{\rm inst}\, ,
\qquad
\tzeta_0 =\, \cl0-\frac{1}{6}\, \kappa_{abc} \,b^a b^b (c^c-\tau_1 b^c)+\tzeta_0^{\rm inst}\, ,
\\
\sigma &= -2 (\psi+\frac12  \tau_1 \cl0) + \cla (c^a - \tau_1 b^a)
-\frac{1}{6}\,\kappa_{abc} \, b^a c^b (c^c - \tau_1 b^c)+\sigma^{\rm inst}\, .
\end{split}
\ee
The relation between $r=e^\phi$ and $\tau_2$ can be obtained by evaluating the contact potential.
In our case, the formula \eqref{solcontpot} gives
\be
e^\phi=\frac{ \tau_2^2}{2} \, \cV
-\frac{\tau_2}{16\pi}{\sum_{m,n}}'
\oint_{C_{m,n}}\frac{\de\varpi}{\varpi}
\(\varpi^{-1} z^{\Lambda}-\varpi \bz^{\Lambda} \)
\p_{\xi^\Lambda}G_{m,n}^{\rm D1}.
\label{contpIIB}
\ee
Again using \eqref{transderGmn}, one may check that the integrand times $\sqrt{\tau_2}$
is invariant under $SL(2,\IZ)$ transformations up to a total derivative.
As a result, the contact potential \eqref{contpIIB} transforms exactly as required in \eqref{SL2phi}.

The formulas \eqref{qmirmapb} and \eqref{contpIIB} encode in a nice and compact way
the quantum corrections to the mirror map. They can be shown to be consistent with
previously known results.
Indeed, evaluating the integrals by residues, one can verify that \eqref{contpIIB}
coincides with the contact potential found in
\cite{RoblesLlana:2006is,Alexandrov:2009qq}, while
the quantum corrections \eqref{qmirmapb} reproduce the results
in Eq. (5.2) \cite{Alexandrov:2009qq} up to
\be
\delta \tzeta_0^{\rm inst} =\sum_{q_a \ge 0} n_{q_a}^{(0)}{\sum\limits_{m,n}}' m\delta_{m,n},
\qquad
\delta\sigma^{\rm inst} =-\sum_{q_a \ge 0} n_{q_a}^{(0)}{\sum\limits_{m,n}}' (m\tau_1+2n)\delta_{m,n},
\label{adcontr}
\ee
where
\be
\delta_{m,n}=\frac{\I n_{q_a}^{(0)}}{16\pi^3}\,
\frac{\tau_2^2}{|m\tau+n|^4} \(1+2\pi q_a t^a |m\tau+n|\) \e^{-S_{m,n,q_a}}.
\ee
Nevertheless, these contributions are harmless since the corresponding corrections to the Darboux coordinates,
which can be written as
\be
\begin{pmatrix}
\delta \txi'_0 \\
\delta\alpha'
\end{pmatrix}
=\sum_{q_a \ge 0} n_{q_a}^{(0)}{\sum\limits_{m,n}}'
\begin{pmatrix} m \\ n \end{pmatrix}
\delta_{m,n}
\label{corr-dc},
\ee
transform as a doublet and preserve the transformation properties \eqref{SL2Zxi}.
Thus, the additional contributions \eqref{adcontr} represent an
inherent ambiguity of the quantum mirror map
and can be dropped at will.

\subsection{Equivalence between IIA and IIB constructions\label{sec_equiv}}

The type IIB construction of the twistor space $\cZ$ outlined above
is manifestly invariant under S-duality, however its equivalence to
the type IIA construction described in Section \ref{subsec-twA} is not manifest.
In  Appendix A.2 of \cite{Alexandrov:2009qq}, it was shown that the type IIA Darboux coordinates
differ from their type IIB counterpart by a complex contact transformation
\be\label{twigauge}
\txi^{[i]}_\Lambda \mapsto \txi^{[i]}_\Lambda - \p_{\xi^\Lambda} H^{[i]}\, ,
\qquad
\ai{i} \mapsto \ai{i} - H^{[i]}+\xi^\Lambda\p_{\xi^\Lambda}H^{[i]} \, .
\ee
In each of the  four quadrants on $\CP$ (in the $\varpi$ coordinate),
away from the real and imaginary axes, this transformation is generated by
\be
H^{[{\rm I}]}=G^{\rm D1}_-,
\qquad
H^{[{\rm II}]}=-G^{\rm D1}_+,
\qquad
H^{[{\rm III}]}=\bG^{\rm D1}_-,
\qquad
H^{[{\rm IV}]}=-\bG^{\rm D1}_+,
\label{gaugefunq}
\ee
where we defined
\be
G^{\rm D1}_\pm(\xi^\Lambda)=
\frac{1}{(2\pi)^2}
{\sum\limits_{
\substack{ (q_a,q_0)\\ \ q_0+q_a b^a>0,\ \pm q_a t^a\ge 0}
}}\, n_{q_a} ^{(0)}\,
\Li_2\left(e^{-2\pi \I q_\Lambda \xi^\Lambda } \right)
+\frac{\chi_\CY}{96}\, ,
\label{prepHpm}
\ee
More succinctly, in all quadrants \eqref{prepHpm} may be written as
\be
\label{contAB}
H=
\sgn\left[ \Re(\varpi)\, \Im(\varpi) \right]\,
\left( \frac{1}{(2\pi)^2}\,
\sum_{
\substack{q_\Lambda\in \IZ\\
(q_0+q_a b^a)\Im\varpi >0,\
 (t^a q_a)\Re(\varpi)\le 0}}
 {\rm Li}_2\left( e^{-2\pi \I q_\Lambda \xi^\Lambda } \right) + \frac{\chi_\CY}{96} \right).
\ee

To understand the origin of this complex contact transformation, it is useful to recall
the structure of the BPS rays $\ell_\gamma$ in the D1-D(-1) sector. These rays extend
from the north to the south pole along the meridian
\be
\arg t = \frac{\pi}{2} + \arg\left[ (q_0+q_a b^a) + \I q_a t^a \right] .
\ee
The conditions $(q_0+q_a b^a)\Im\varpi >0,\ (t^a q_a)\Re(\varpi)<0$ select precisely
those BPS rays which pass through the quadrant where $\varpi$ is located. Thus, the
complex contact transformation generated by \eqref{contAB} is the product of the
elementary contact transformations \eqref{elemks}
associated with each of the BPS rays encountered in
interpolating from a point $\varpi$ in the vicinity of the real axis to the point $t=\I$
if $\Im\varpi>0$, or to the point $t=-\I$ if $\Im\varpi<0$. One can therefore interpret
the type IIB Darboux coordinates as the standard Darboux coordinates in the angular
sector containing the S-duality invariant points $\varpi=\pm \I$, whereas the type IIA
Darboux coordinates are the standard Darboux coordinates in the angular
sectors containing the (positive or negative) real axis. This is consistent with the
fact that the initial step in the computation of  \cite{Alexandrov:2009qq} was to move
all the BPS rays along the positive or negative imaginary axis, before performing
the Poisson resummation over the D(-1)-instanton charge $q_0$. Had one chosen
instead to move all BPS rays along the positive or negative real axis, such a contact
transformation would not have arisen.

\section{S-duality in twistor space with two isometries\label{sec_thm}}

In this section we propose a twistorial construction of a general class of
\qk metrics, which consists of non-linear deformations of the $c$-map metric
\eqref{hypmettree} which preserve two translational isometries along
the coordinates $\psi$ and $\cl0$ and carry the isometric action \eqref{SL2Z}
of $SL(2,\IZ)$. This class of metrics includes the D1-D(-1)-instanton
corrected hypermultiplet metric, and generalizes the
type IIB twistorial construction  described in Section \ref{sec_d1}.
It is tailored for describing D3-instanton corrections in a manifestly
S-duality invariant fashion, with the coordinates $\psi$ and $\cl0$
playing the role of the axions dual to the D5 and NS5-instantons.
However, the identification of the D3-instanton
corrected HM metric within this class goes beyond the scope of this work.

Since the continuous isometries along $\psi$ and $\cl0$
lift to translations along the complex Darboux coordinates $\alpha$ and $\txi_0$, the
holomorphic transition functions appearing in this construction are allowed
to depend on $\xi^0, \xi^a$ and $\txi_a$ in an arbitrary fashion (except
for some global constraints following from S-duality). As a result, the Darboux
coordinate $\xi^0$ is (projectively) globally well defined as the moment
map of the Killing vector $\pa_{\cl0}$, and given in all patches by the
same $\cO(2)$ section as in \eqref{gentwi}.  In contrast, the Darboux coordinates
$\xi^a, \txi_\Lambda,\alpha$ are defined only locally.

Let  $\cZ$ be the $2n+1$-complex dimensional contact manifold defined by the infinite covering
\be
\cZ= \cU_+\cup\cU_- \cup\cU_{0}\cup\(\cup_{m,n}' \cU_{m,n}\) ,
\ee
and transition functions
\be
\label{transfun-gen}
\Hij{+0}=\Fcl(\xii{+}^\Lambda) ,
\qquad
\Hij{-0}=\bFcl(\xii{-}^\Lambda) ,
\qquad
\Hij{(m,n)0}= G_{m,n}(\xi^0,\xii{m,n}^a,\txii{0}_a)\ ,
\ee
where $\Fcl(X)=-\kappa_{abc} \,\frac{X^a X^b X^c}{6 X^0}$ is an arbitrary cubic prepotential,
$a=1,\dots n-1$, $\Lambda=0,1,\dots {n-1}$.
Here $\cU_\pm$ are the usual patches around the poles of $\CP$,
$\cU_{m,n}$ is a set of patches which are mapped to each other under $SL(2,\IZ)$-transformations,
and $\cU_0$ covers the rest.
They must be chosen so that $\cU_0$ is mapped to itself under the antipodal map
and S-duality, whereas $\cU_{m,n}$ are mapped to each other according to
\be
\tau\[\cU_{m,n}\]= \cU_{-m,-n},
\qquad
\cU_{m,n}\mapsto \cU_{m',n'},
\qquad
\( m'\atop n'\) =
\(
\begin{array}{cc}
a & c
\\
b & d
\end{array}
\)
\( m \atop n \).
\ee
The local holomorphic functions $G_{m,n}$ are assumed to transform under the $SL(2,\IZ)$
action \eqref{SL2Zxi}
as
\be
G_{m,n}
\mapsto \frac{G_{m',n'}}{c\xi^0+d}
-\frac{c}{2}\, \frac{\kappa_{abc}\p_{\txii{0}_a}G_{m',n'}\p_{\txii{0}_b}G_{m',n'}}{(c\xi^0+d)^2}
\(\xii{m',n'}^c-\frac{2}{3}\,\p_{\txii{0}_c}G_{m',n'}\)
+\mbox{reg.}
\label{StransG}
\ee
where $+ \mbox{reg.}$  denotes equality up to terms which are regular in $\cU_{m',n'}$.
Then the following statements hold:
\begin{enumerate}
\item
The Darboux coordinates in the patch $\cU_0$ satisfy the following integral
equations:
\bea
\hspace{-1cm}
\xi^0 &=& \zeta^0+\frac{\tau_2}{2}\(\varpi^{-1}-\varpi\),
\nn
\\
\hspace{-1cm}
\xii{0}^a &=& \zeta_{\rm cl}^a +\varpi^{-1}Y^a -\varpi \bY^a+
 {\sum_{m,n}}' \oint_{C_{m,n}} \frac{\de \varpi'}{2 \pi \I \varpi'} \,
K(\varpi,\varpi') \, \p_{\txii{0}_a}G_{m,n},
\nn
\\
\hspace{-1cm}
\txii{0}_\Lambda& =& \tzeta^{\rm cl}_\Lambda +\varpi^{-1}\Fcl_\Lambda(Y)-\varpi\bFcl_\Lambda(\bY)
- {\sum_{m,n}}' \oint_{C_{m,n}} \frac{\de \varpi'}{2 \pi \I \varpi'} \,
K(\varpi,\varpi')\, \p_{\xii{m,n}^\Lambda}G_{m,n},
\label{twistlines}
\\
\hspace{-1cm}
\ai{0}&= &  -\hf(\tsigma+\zeta^\Lambda \tzeta_\Lambda)^{\rm cl}
-\(\varpi^{-1}+\varpi\)\(\varpi^{-1} \Fcl(Y)+\varpi \bFcl(\bY)\)
-\zeta_{\rm cl}^\Lambda\( \varpi^{-1}\Fcl_\Lambda(Y)-\varpi\bFcl_\Lambda(\bY)\)
\nn \\
&-&
{\sum_{m,n}}' \oint_{C_{m,n}} \frac{\de \varpi'}{2 \pi \I \varpi'} \,
\[ K(\varpi,\varpi')
 \(1-\xii{m,n}^\Lambda(\varpi') \p_{\xii{m,n}^\Lambda}\)G_{m,n}
+\frac{ (\varpi \varpi')^{-1} \Fcl_a(Y)+\varpi \varpi'\bFcl_a(\bY) }{1/\varpi'+\varpi'}\,\p_{\txii{0}_a}G_{m,n} \],
\nn
\eea
where $C_{m,n}$ are contours on $\CP$ surrounding $\cU_{m,n}$ counter-clockwise
and $(\tau_2,Y^a,\zeta^0,\zeta_{\rm cl}^a,\tzeta^{\rm cl}_a,\tsigma^{\rm cl})$
are coordinates on $\cM$.

\item  Upon identifying these coordinates with the following combinations of the
type IIB fields $(\tau_1,\tau_2,t^a,b^a,c^a,\cla,\cl0,\psi)$,
\be
\label{mirror-map}
\begin{split}
\zeta^0 =& \tau_1,
\qquad
\zeta_{\rm cl}^a = - (c^a - \tau_1 b^a),
\\
\tzeta^{\rm cl}_a =&  \cla+ \frac{1}{2}\, \kappa_{abc} \,b^b (c^c - \tau_1 b^c) ,
\qquad
\tzeta^{\rm cl}_0 =\, \cl0-\frac{1}{6}\, \kappa_{abc} \,b^a b^b (c^c-\tau_1 b^c),
\\
\tsigma^{\rm cl} =& -2 (\psi+\frac12  \tau_1 \cl0) + \cla (c^a - \tau_1 b^a)
-\frac{1}{6}\,\kappa_{abc} \, b^a c^b (c^c - \tau_1 b^c)
+\frac{\tau_2^2}{3}\, \kappa_{abc}b^a b^b b^c,
\\
Y^0 =& \frac{\tau_2}{2},
\qquad
Y^a =\frac{\tau_2}{2}\( b^a+\I t^a\)+{\sum_{m,n}}'
\oint_{C_{m,n}} \frac{\de \varpi}{2 \pi \I \varpi^2} \,
\frac{\p_{\txii{0}_a}G_{m,n}}{\(1/\varpi+\varpi\)^2},
\end{split}
\ee
the Darboux coordinates \eqref{twistlines} transform under
the $SL(2,\IZ)$ transformations \eqref{SL2Z} and \eqref{SL2varpi}
according to the classical laws \eqref{SL2Zxi}.

\item The same transformation rules also hold for the Darboux coordinates
in the patches $\cU_{m,n}$, with the understanding that the Darboux coordinates
appearing in the r.h.s. are those attached to the patch $\cU_{m',n'}$,
e.g. $\xi^a_{[m,n]} \mapsto \xi^a_{[m',n']}/(c\xi^0+d)$.

\item
The contact potential is given by
\bea
\label{fullephi}
e^{\Phi}& =&\frac{2}{3\tau_2}\,\kappa_{abc} \Im Y^a\Im Y^b \Im Y^c
\\
&& -\frac{1}{16\pi}{\sum_{m,n}}'
\oint_{C_{m,n}}\frac{\de\varpi}{\varpi}\[
\(\varpi^{-1} Y^{\Lambda}-\varpi \bY^{\Lambda} \)\p_{\xii{m,n}^\Lambda}G_{m,n}
+\(\varpi^{-1} \Fcl_{a}(Y)-\varpi \bFcl_a(\bY) \)\p_{\txii{0}_a}G_{m,n}\]
\nn
\eea
and transforms under $SL(2,\IZ)$ as in \eqref{SL2phi}.

\item As a result, the $4n$-dimensional \qk manifold $\cM$ associated to the twistor space $\cZ$
carries an isometric action of $SL(2,\IZ)$.

\item Upon freezing the moduli $\tau_1,\tau_2$ and ignoring the Darboux
coordinates $\alpha,\txi_0$, the remaining ones, $\xi^a$ and $\txi_a$, determine
a one-parameter family of $(4n-4)$-dimensional \hk metrics $\cM_{\tau}$,  with $\tau$ taking
values in the Poincar\'e upper half plane, such that $\cM_{\tau}$ and $\cM_{(a\tau+b)/(c\tau+d)}$
are identified isometrically under \eqref{SL2Z}.

\end{enumerate}

Some comments are in order:
\begin{itemize}

\item
The key transformation law \eqref{StransG}
expresses the fact that S-duality should commute with the holomorphic
transition functions up to local contact transformations, i.e. that
the following diagram should be commutative:
\be
\begin{array}{rcl}
\vphantom{ \frac{A}{\displaystyle A_A}}
\cU_0\ \ & \stackrel{\displaystyle S}{\longrightarrow}\ &\ \cU_0 \\
\vphantom{A^A\over A}
G_{m,n} \ \downarrow\ \ \ & &\ \ \downarrow \ G_{m',n'}\\
\cU_{m,n} & \stackrel{\displaystyle S}{\longrightarrow} \ & \cU_{m',n'}
\end{array}
\ee
Here the horizontal arrows correspond to the holomorphic action \eqref{SL2Zxi} of S-duality
on twistor space, which maps the Darboux coordinates in $\cU_0$ (resp., $\cU_{m,n}$) to
Darboux coordinates in the same patch $\cU_0$ (resp., in the transformed patch $\cU_{m',n'}$).
The quadratic and cubic terms in \eqref{StransG} have the same origin as the
quadratic term in \eqref{transellg}, namely  the fact that the generating function
of the contact transformation must be expressed in terms of the coordinates $\xi^{\Lambda}$ in the
original patch and of the coordinates $\txi_\Lambda$ in the final patch. In principle, the condition
\eqref{StransG} determines all $G_{m,n}$ in terms of $G_{({\rm gcd}(m,n),0)}$,
although the non-linearities in \eqref{StransG} prevent us from writing a closed expression.
In the one-instanton approximation, the quadratic and cubic terms can be omitted,
and the transformation law \eqref{StransG} simply expresses the invariance of a Cech
cocycle in $\cH^1(\cZ,\cO(2))$
under $SL(2,\IZ)$. It may be summarized by saying that
the formal sum $\sum'_{m,n} G_{m,n}$  transforms as a holomorphic modular form
of weight $-1$.

\item The equations \eqref{twistlines} are simply the translation of
the general integral equations \eqref{txiqline} for the specific choice of contours
and transition functions \eqref{transfun-gen}, after performing the change of kernel
\eqref{modker}. In particular the last term in the square bracket in the expression for
$\ai{0}$ originates from the transition
functions $\Hij{\pm 0}$ and may be rewritten as $[t^{-1} K(0,t')F^{\rm cl}_a
-t K(\infty,t') \bar F^{\rm cl}_a]\p_{\txii{0}_a}G_{m,n}$. The kernel substitution
does not affect the discontinuity along the contours,
and can be reabsorbed by the change of coordinates \eqref{mirror-map}.

\item If $G_{m,n}$ is independent of $\txii{0}_a$, we recover the construction
of the D1-D(-1)-instanton corrected metric in Section \ref{sec_d1}. In that case
the patches $\cU_{m,n}$ were centered around the zeros of $m\xi^0+n$,
which we denoted by $\tzpmcd{m,n}$, and the transition functions are given in \eqref{defG1}.
Their transformation law \eqref{transGmn} is a particular case of \eqref{StransG}.

\item
Instead or in addition to the closed contours $C_{m,n}$ surrounding open patches $\cU_{m,n}$,
one can include open contours similar to the BPS rays appearing in
the construction of Section \ref{subsec-twA}.
In fact, all previous statements hold for an arbitrary set of contours
$C_{m,n}$ mapped unto each other by $SL(2,\IZ)$.
For example, $C_{m,n}$ can be an open contour from $\tzpcd{m,n}$ to $\tzmcd{m,n}$.
In this case however the structure of patches might be more complicated.
The only requirement on the covering is that
it should be invariant under the antipodal map and $SL(2,\IZ)$ transformations so that
a patch $\cU_i$ is mapped to some patch $\cU_{i'}$. The Darboux coordinates in each patch
are given by the formulas \eqref{twistlines} and transform under S-duality
by \eqref{SL2Zxi} where the coordinates on the r.h.s. are from the patch $\cU_{i'}$.
Moreover, if the function $G_{m,n}$ is associated with an open contour, the regular terms
in the transformation rule \eqref{StransG} are no longer allowed, as they would otherwise
not disappear under integration and therefore spoil S-duality.

\item
Note that the coordinate change \eqref{mirror-map} is almost identical to
the classical mirror map \eqref{symptobd}, up to two differences.
First, the coordinate $\tilde\sigma^{\rm cl}$ differs from $\sigma$ in \eqref{symptobd}
by the last term $\frac{\tau_2^3}{3}\kappa_{abc} b^a b^b b^c$. Although this
is a classical contribution, we chose to absorb it into $\tsigma^{\rm cl}$ in order
to simplify the expression for $\alpha$ in terms of $Y^a$.
Second and more importantly, $Y^a$ receives instanton corrections away from its classical
value $Y^a=\frac{\tau_2}{2}\, z^a$. This may be viewed as an instanton correction to the mirror map,
however note that the notion of mirror map only makes sense for the specific choice of
$G_{m,n}$ corresponding to D3-brane instantons. In that case, the full quantum mirror map
 will require specifying also $\zeta^{a}, \tzeta_\Lambda, \sigma$ in terms of
$\zeta_{\rm cl}^a,\tzeta^{\rm cl}_a,\tsigma^{\rm cl}$.

\item
The identifications \eqref{mirror-map} are not the only possible ones which lead to
manifest $SL(2,\IZ)$ invariance. We already met an example of such an ambiguity in \eqref{adcontr}
when studying D1-D(-1)-instanton corrections to the mirror map.
Another simple example can be found in Appendix \ref{ap-check}, Eq. \eqref{addtzeta},
and many other examples can be constructed (see e.g. \cite{amp-to-appear}).
Fortunately, all these ambiguities just represent a freedom in the choice
of coordinates and do not affect the geometry of the QK manifold.

\item The modular invariant family of $4n-4$-dimensional HK manifolds
$\cM_{\tau}$ arises as a rigid limit of the QK space $\cM$, using the same philosophy as
in the QK/HK correspondence studied in \cite{Alexandrov:2011ac}. Namely, the two isometries
$\pa_{\cl0}$ and $\pa_{\psi}$ on $\cM$ can be lifted to triholomorphic isometries of the
Swann bundle $\cS$ (a $\IC^\times$ bundle over $\cZ$), with moment maps $\eta^\flat$
and $\eta^0 = \xi^0 \eta^\flat$ in the notations of \cite{Alexandrov:2008nk}. The \hk
quotient of $\cS$ with respect to these two isometries is obtained by enforcing the
D-term constraints, i.e. freezing the
$\cO(2)$ multiplets $\eta^\flat$
and $\eta^0$, and quotienting by the action of $\pa_{\cl0}$ and $\pa_{\psi}$. It
produces a family of $(4n-4)$-dimensional
\hk  metrics, parametrized by the moment maps $\eta^\flat$ and $\eta^0$.
The former can be fixed to an arbitrary non-zero value by a suitable $SU(2)$ rotation, while the latter
yields one complex parameter $\tau$ which is extracted from the components of the $\cO(2)$ multiplet $\xi^0$.
For fixed value of $\tau$, the HK metric is then coordinatized by $t^a,b^a,c^a,\cla$,
or by the Darboux coordinates $\xi^a,\txi_a$. This class of modular invariant HK
metrics should include the Coulomb branch of five-dimensional $\cN=2$ gauge theories
compactified on $T^2$, where the modular group of $T^2$ plays the role of
S-duality \cite{Haghighat:2011xx} and the monopole strings play the role of D3-instantons.

\end{itemize}

We emphasize that the statements (1.-6.) hold to all orders in the instanton expansion.
It is a tedious, but straightforward exercise to verify them.
To this end, one first brings the integral equations \eqref{txiqline} to the form \eqref{twistlines}
and the contact potential \eqref{solcontpot} to the form \eqref{fullephi}.
This can be done by evaluating the integrals with $\Hij{\pm0}$ and by a suitable identification
of $A^\Lambda,B_\Lambda,B_\alpha$ with $\zeta_{\rm cl}^\Lambda,\tzeta^{\rm cl}_\Lambda,\tsigma^{\rm cl}$.
Then $SL(2,\IZ)$ transformations of the Darboux coordinates and
the contact potential can be checked by using the
transformation properties of the derivatives of $G_{m,n}$, which provide a generalization
of \eqref{transderGmn} and can be easily obtained from \eqref{StransG},
\bea
\hspace{-1cm}
\p_{\txii{0}_a}G_{m,n} &\ \mapsto\ & \frac{\p_{\txii{0}_a}G_{m',n'}}{c\xi^0+d}
+\mbox{reg.}
\nn
\\
\hspace{-1cm}
\p_{\xii{m,n}^a}G_{m,n} &\ \mapsto\ & \p_{\xii{m',n'}^a}G_{m',n'}
-\frac{c\,\kappa_{abc}\p_{\txii{0}_b}G_{m',n'}}{c\xi^0+d}\(\xii{m',n'}^c-\hf\,\p_{\txii{0}_c}G_{m',n'}\)
+\mbox{reg.}
\nn
\\
\hspace{-1cm}
\p_{\xi^0}G_{m,n} &\ \mapsto\ & d\, \p_{\xi^0}G_{m',n'}-c\, (1-\xii{m',n'}^\Lambda\p_{\xii{m',n'}^\Lambda})G_{m',n'}
\label{StransderG}
\\
\hspace{-1cm}
&&-
\frac{c^2 \kappa_{abc}\p_{\txii{0}_a}G_{m',n'}}{2(c\xi^0+d)}\(\xii{m',n'}^b \xii{m',n'}^c
-\xii{m',n'}^b \p_{\txii{0}_c}G_{m',n'}+\frac{1}{3}\,\p_{\txii{0}_b}G_{m',n'}\p_{\txii{0}_c}G_{m',n'}\)\nn\\
&& +\mbox{reg.}
\nn
\eea
In Appendix \ref{ap-check}, we explain in detail how
the construction works
in the one-instanton approximation, i.e. to first order in $G_{m,n}$.

\section{Discussion}
\label{sec-concl}

In this work we have provided a twistorial construction for a class of \qk metrics which
admit two commuting, continuous isometries and an isometric action of $SL(2,\IZ)$.
These metrics are non-linear deformations of the standard $c$-map metric derived
from a cubic prepotential. They are parametrized by an infinite set of holomorphic functions
$G_{m,n}(\xi^\Lambda,\txi_a)$ subject to the transformation rule \eqref{StransG} and
tailored for describing D3-instanton corrections to the QK
metric on the HM moduli space $\cM_H$ in type IIB string theory compactified on a
Calabi-Yau threefold. As a special case, one recovers the D1-D(-1)-instanton corrected
metric on $\cM_H$ constructed in \cite{RoblesLlana:2006is,Alexandrov:2009qq},
whose manifest S-duality invariance is now revealed. Upon taking a rigid limit,
the same construction also yields a class of modular invariant \hk metrics (where
the modulus $\tau$ is now a parameter rather than a coordinate), which should
describe the Coulomb branch of five-dimensional $\cN=2$ gauge theories
compactified on $T^2$, with monopole strings playing the role of D3-instantons.

Clearly, an outstanding question is to determine the set of holomorphic functions
$G_{m,n}$ which describes D3-instanton corrections (or similarly, the monopole
string instantons) to the real HM moduli space.
For this purpose one should find an adequate complex contact transformation which
would cast the type IIA construction of the twistor space outlined in Section \ref{subsec-twA}
into the manifestly  S-duality invariant type IIB construction of Section \ref{sec_thm}.
This is an important problem, since it would prove that D3-instanton corrections are
indeed consistent with S-duality, and it would allow to compute quantum corrections
to the mirror map.

An indication that this translation can be done is the fact that, in the one-instanton approximation,
the formal sum of transition functions $H_\gamma$ in \eqref{transellg} over all D3-D1-D(-1)
charges is formally identical to the partition function of D4-D2-D0 black holes
in type IIA string theory compactified on the same CY threefold $\CYm$. The latter is a
Jacobi form and admits a natural Poincar\'e series representation
\cite{Maldacena:1997de,Gaiotto:2006ns, Gaiotto:2006wm,deBoer:2006vg,Denef:2007vg}.
Unfortunately, there are technical difficulties in implementing this idea due to the fact that,
for fixed D3-brane charge, the formal sum over D1-instantons leads to an indefinite theta series of
signature $(1,b_2(\CYm)-1)$ which is divergent. Moreover, Poincar\'e series
of negative weight are also divergent and need to be regularized, leading
to holomorphic or modular anomalies.
In \cite{amp-to-appear}, we attack this problem from a different angle,
and show that the instanton corrections to the type IIA Darboux coordinates, in the large volume limit,
are governed by certain Mock theta series. In the analogous problem of
the Coulomb branch of $\cN=2$ five dimensional gauge theories,
the monopole string instantons are described by the elliptic
genus of the $(0,4)$ superconformal field theory with target space given by the ADHM
moduli space of monopoles, which is also expected to be modular invariant \cite{Haghighat:2011xx}.
It would be very interesting to see if some of the difficulties of string theory are
alleviated in the gauge theory set-up.

Another outstanding problem is to remove the assumption that the metric has
two commuting Killing vectors, and to construct a general class of $SL(2,\Z)$ invariant
QK metrics with no continuous isometries as non-linear perturbations of the $c$-map
metric. Similar to the construction in Section \ref{sec_thm}, this class should be
parametrized by functions $G_{m,n}$ which are now allowed to depend on all
Darboux coordinates $\xi^\Lambda,\txi_\Lambda,\alpha$. At the linear level,
S-duality is again ensured by the condition \eqref{StransG} where
only the first term on the r.h.s. should be retained. An example of
such construction was given in \cite{Alexandrov:2010ca}, where NS5-brane
instanton corrections were inferred by S-duality from the D5-brane instantons
and shown to be governed by a set of transition functions transforming with modular
weight $-1$ as in \eqref{StransG}.
However, an extension of this construction beyond the linear level appears to be
highly non-trivial.

\acknowledgments

We are grateful to D. Persson and S. Vandoren for valuable discussions,
and to J. Manschot for discussions and collaboration on a closely related project.

\appendix

\section{Twistorial description of \qk manifolds \label{secQK}}

 A $4n$-dimensional Riemannian manifold $\cM$ is \qk if it has restricted
holonomy group $USp(n)\times SU(2) \subset SO(4n)$. The Ricci scalar $R$ is then
constant, and the curvature of the $SU(2)$ part of the Levi-Civita connection, rescaled
by $1/R$, provides a  triplet of quaternionic 2-forms $\vec \omega$. While $R$ can
take either sign, hypermultiplet moduli spaces  in $\cN=2$ supergravity or string theory
models have $R<0$.

A QK manifold $\cM$ can be described analytically in terms of its twistor space $\cZ$,
the total space  of the $\CP$ bundle over
$\cM$ twisted with the \emph{projectivized} $SU(2)$ connection on $\cM$.
$\cZ$ is a \kahler-Einstein space equipped with a canonical complex contact structure,
given by the kernel of the one-form
\be
\label{defDt}
D\varpiqk= \de \varpiqk+p_+ - \I p_3  \varpiqk+p_- \varpiqk^2,
\ee
where $\varpiqk$ is a complex coordinate on $\CP$, and $p_\pm,p_3$
is the $SU(2)$ part of the Levi-Civita connection on $\cM$.
Note that $D\varpiqk$ is defined only projectively, as it rescales under $SU(2)$
rotations. More precisely, it is valued in the $\cO(2)$ line bundle on $\CP$ \cite{MR664330}.

Locally, on a patch
of an open covering $\{\hcU_i\}$ of $\cZ$,
one can always find complex  Darboux coordinates  $(\xii{i}^\Lambda,\txii{i}_\Lambda,\ai{i})$
such that the contact one-form \eqref{defDt}, suitably rescaled by a function $e^{\Phi\ui{i}}$, takes the form
\be
\cX^{[i]}\equiv 4 \, e^{\Phi\ui{i}}\, \frac{D\varpiqk}{\I\varpiqk} =  \de\ai{i}+ \txii{i}_\Lambda \de \xii{i}^\Lambda\, .
\label{contform}
\ee
The function $\Phi\ui{i}$, which we refer to as the `contact potential', is holomorphic along
the $\CP$ fiber and defined up to an additive holomorphic function on
$\hcU_i$.  It provides
a K\"ahler potential for the \kahler-Einstein metric on $\cZ$ \cite{Alexandrov:2008nk}:
\be
\label{Knuflat}
K\ui{i}_{\cZ} = \log\frac{1+\varpiqk\bar \varpiqk}{|\varpiqk|}+\Re\Phi\ui{i}\, .
\ee

Globally, the complex contact structure on $\cZ$ can be specified by a set of generating
functions $\qHij{ij}(\xii{i}^\Lambda,\txii{j}_\Lambda,\ai{j})$ for complex contact transformations
between Darboux coordinates on overlaps
$\hcU_i\cap \hcU_j$,
subject to cocycle and reality conditions \cite{Alexandrov:2008nk}.

In the case when $\cM$ has a quaternionic isometry
$\partial_\sigma$, one may choose the Darboux coordinates such that the Killing
vector lifts to the holomorphic action $\pa_\alpha$. As a result, the
transition functions $\qHij{ij}$ become independent of the coordinate $\alpha\ui{j}$,
and the contact potential $\Phi^{[i]}$ becomes real and constant on
$\CP$ \cite{Alexandrov:2008gh}, equal to log-norm of the moment map associated to the
isometry. The twistorial construction of such QK
manifolds with one quaternionic isometry then becomes isomorphic to the twistorial
construction of HK manifolds with one rotational isometry, a relation known as the
QK/HK correspondence \cite{Haydys,Alexandrov:2011ac}.
In this case the Darboux coordinates are determined by the following system
of integral equations:
\beq
\xii{i}^\Lambda(\varpiqk)& =& A^\Lambda +
\varpiqk^{-1} Y^\Lambda - \varpiqk \, \bY^\Lambda
-\frac12 \sum_j \int_{C_j}\frac{\de\varpiqk'}{2\pi\I \varpiqk'}\,
\frac{\varpiqk'+\varpiqk}{\varpiqk'-\varpiqk}
\,\p_{\txii{j}_\Lambda }\qHij{ij},
\nonumber \\
\txi_\Lambda^{[i]}(\varpiqk)& = &  B_\Lambda +
\half  \sum_j \int_{C_j} \frac{\de \varpiqk'}{2 \pi \I \varpiqk'} \,
\frac{\varpiqk' + \varpiqk}{\varpiqk' - \varpiqk}
\, \p_{\xii{i}^\Lambda } \qHij{ij},
\label{txiqline}
\\
\ai{i}(\varpiqk)& = & B_\alpha +
\half  \sum_j \int_{C_j} \frac{\de \varpiqk'}{2 \pi \I \varpiqk'} \,
\frac{\varpiqk' + \varpiqk}{\varpiqk' - \varpiqk}
\( \qHij{ij}- \xii{i}^\Lambda \p_{\xii{i}^\Lambda}\qHij{ij}\)
+4\I c \log\varpiqk .
\nonumber
\eeq
Here the complex variables $Y^\Lambda$, up to an overall phase rotation
which can be absorbed into a phase rotation of $\varpiqk$,
and the real variables $A^\Lambda,B_\Lambda,B_\alpha$
serve as coordinates on $\cM$. It is convenient to fix the phase freedom in  $Y^\Lambda$
by requiring $Y^0\equiv \cR$ to be real.
Moreover, $B_\alpha$ is related to the coordinate $\sigma$ along the isometric direction,
$\pa_{B_\alpha}\sim \pa_\sigma$. Finally,
$c$ is a real constant known as the anomalous dimension \cite{Alexandrov:2008nk},
which characterizes the singular behavior of the Darboux coordinate $\alpha$
at the north and south poles of $\CP$. It plays an important physical role in describing
the one-loop correction to the hypermultiplet moduli space metric in type II
string compactifications.

The integrals in \eqref{txiqline} are taken around closed contours $C_i$ surrounding
the patches $\cU_i$ in the counter-clockwise direction. Nevertheless, the  construction
is still meaningful if some of the contours are taken to be open. Typically such open contours
appear as the degeneration of closed contours in the presence of branch cut singularities
in the transition functions $\Hij{ij}$. The holomorphic functions associated with that
open contours are then equal to the discontinuity of  $\Hij{ij}$ across the branch cut \cite{Alexandrov:2008ds}.
In particular, this is the case for twistorial constructions of D-brane instantons
in type IIA \cite{Alexandrov:2008gh,Alexandrov:2009zh} as well as fivebrane
instantons in type IIB \cite{Alexandrov:2010ca}.

Thus, the metric on $\cM$ is completely determined by
a constant $c$ and a set of open or closed contours in $\CP$, together with associated holomorphic functions.
The procedure to extract the metric from the solutions of \eqref{txiqline} was outlined
in \cite{Alexandrov:2008nk,Alexandrov:2008gh}.
An important ingredient in this calculation is the contact potential, which can be computed from the transition
functions $\qHij{ij}$ and the solutions of  \eqref{txiqline} using
\be
e^{\Phi}=\frac{1}{16\pi} \sum_j\int_{C_j}\frac{\de\varpiqk}{\varpiqk}
\(\varpiqk^{-1} Y^{\Lambda}-\varpiqk \, \bY^{\Lambda} \)
\p_{\xii{i}^\Lambda } \qHij{ij}
-c.
\label{solcontpot}
\ee

\section{S-duality invariance at linear order}
\label{ap-check}

In this Appendix  we show the consistency of the construction of Section \ref{sec_thm}
in the one-instanton approximation. Our aim is to prove that the Darboux coordinates
\eqref{twistlines} transform under S-duality as in \eqref{SL2Zxi}, provided the
coordinates $(\tau_2,Y^a,\zeta^0,\zeta_{\rm cl}^a,\tzeta^{\rm cl}_a,\tsigma^{\rm cl})$
are related to the type IIB fields $(\tau_1,\tau_2,t^a,b^a,c^a,\cla,\cl0,\psi)$ as in \eqref{mirror-map}.

We start with the Darboux coordinate $\xii{0}^a$. Under an $SL(2,\IZ)$ transformation,
the kernel $K(\varpi,\varpi') \de \varpi'/\varpi'$ is invariant but the derivative
$\p_{\txii{0}_a}G_{m,n}(t')$
transforms with an overall factor of $1/(c \xi^0(\varpi')+d)$ (see \eqref{StransderG}).
Remarking that $t^{-1}+t$ transforms as
\be
\label{transtti}
\varpi^{-1}+\varpi \ \mapsto\  \frac{|c\tau+d|}{c\xi^0+d}\(\varpi^{-1}+\varpi\) ,
\ee
we see that the factor  $1/(c \xi^0(\varpi')+d)$ can be converted into $1/(c \xi^0(\varpi)+d)$
by inserting an additional factor of $(t^{-1}+t)/(t'^{-1}+t')$ in the  integrand. This is the purpose
of the field redefinition of $Y^a$ in \eqref{mirror-map}. Indeed,
by  expressing $Y^a$ in terms of $z^a=b^a+\I t^a$ in the second line of
\eqref{twistlines} we arrive at
 \be
\xii{0}^a = \zeta_{\rm cl}^a +\frac{\tau_2}{2}\(\varpi^{-1}z^a -\varpi \bz^a\)+ \delta\xi^a,
\ee
where
\be
\delta\xi^a =
 {\sum_{m,n}}' \oint_{C_{m,n}} \frac{\de \varpi'}{2 \pi \I \varpi'} \,
\frac{1/\varpi+\varpi}{1/\varpi'+\varpi'}\,  K(\varpi,\varpi')
\, \p_{\txi_a}G_{m,n}.
\label{twlinexianew}
\ee
Both terms then manifestly transform as in \eqref{SL2Zxi}, under the assumption
that $\zeta_{\rm cl}^a$ is related to $c^a$ as in \eqref{mirror-map}.

To investigate the other Darboux coordinates in \eqref{twistlines}, we first  rewrite
them in terms of the type IIB \kahler moduli $z^a$ as
\be
\begin{split}
\txii{0}_a =&\, \tzeta^{\rm cl}_a -\frac{\tau_2}{4}\,
\kappa_{abc}\(\varpi^{-1}z^b z^c-\varpi\bz^b\bz^c\) + \delta\txi_a,
\\
\delta\txi_a & \,=
- {\sum_{m,n}}' \oint_{C_{m,n}} \frac{\de \varpi'}{2 \pi \I \varpi'} \,
\( K(\varpi,\varpi')\, \p_{\xi^a}G_{m,n}
+\kappa_{abc}\frac{(\varpi\varpi')^{-1} z^b -\varpi\varpi'\bz^b}{(1/\varpi'+\varpi')^2}\, \p_{\txi_c}G_{m,n}
\),
\\
\txii{0}_0 =&\, \tzeta^{\rm cl}_0 +\frac{\tau_2}{12}\,\kappa_{abc}\(\varpi^{-1}z^a z^b z^c-\varpi\bz^a \bz^b\bz^c\) + \delta\txi_0,
\label{twlineIIBtxi}
\\
\delta\txi_0 &\,= - {\sum_{m,n}}' \oint_{C_{m,n}} \frac{\de \varpi'}{2 \pi \I \varpi'} \,
\( K(\varpi,\varpi')\, \p_{\xi^0}G_{m,n}
- \kappa_{abc}\frac{(\varpi\varpi')^{-1}z^a z^b-\varpi\varpi'\bz^a\bz^b}{2(1/\varpi'+\varpi')^2}\, \p_{\txi_c}G_{m,n}
\),
\\
\ai{0}= &\,  -\hf(\sigma+\zeta^\Lambda \tzeta_\Lambda)_{\rm cl}
+\frac{\tau_2^2}{8}\(\bz^\Lambda \Fcl_\Lambda(z)+z^\Lambda \bFcl_\Lambda(\bar z)\)
 \\
&
-\frac{\tau_2^2}{4}\( \varpi^{-2} \Fcl(z)+\varpi^2 \bFcl(\bz)\)-\frac{\tau_2\zeta^\Lambda}{2}
\( \varpi^{-1}\Fcl_\Lambda(z)-\varpi\bFcl_\Lambda(\bz)\) + \delta\alpha,
\\
\delta\alpha&\,=-{\sum_{m,n}}' \oint_{C_{m,n}} \frac{\de \varpi'}{2 \pi \I \varpi'} \,
\( K(\varpi,\varpi')  \(1-\xi^\Lambda(\varpi')\p_{\xi^\Lambda}\)G_{m,n}
-\kappa_{abc}\frac{V^{ab}(\varpi,\varpi')  }{(1/\varpi'+\varpi')^2} \,\p_{\txi_c}G_{m,n}\),
\end{split}
\ee
where, in the last expression for $\delta\alpha$,
\be
\label{contrV}
\begin{split}
V^{ab}(\varpi,\varpi')
=&\, -c^a\( b^b\frac{1-(\varpi\varpi')^2}{\varpi\varpi'}+\I t^b\frac{1+(\varpi\varpi')^2}{\varpi\varpi'}\)
+\frac{\tau_1}{2} \(b^a b^b +t^a t^b\)\frac{1-(\varpi\varpi')^2}{\varpi\varpi'}
\\
&
+\frac{\tau_2}{4}\( \varpi^{-1}+\varpi+\varpi'^{-1}+\varpi'\)
\( \(b^a b^b - t^a t^b\)\frac{1+(\varpi\varpi')^2}{\varpi\varpi'}+2\I b^a t^b\frac{1-(\varpi\varpi')^2}{\varpi\varpi'}\).
\end{split}
\ee
It may be checked by a direct but tedious calculation that each of these Darboux coordinates transforms as in \eqref{SL2Zxi},
provided $\tzeta^{\rm cl}_\Lambda,\sigma_{\rm cl}$ are related to $c_a, c_0,\psi$ as in \eqref{mirror-map}.

This computation is however best performed
by introducing first the following linear combinations of the instanton contributions
\be
\label{defdeltahx}
\begin{split}
\hat\delta\txi_a =&\, \delta\txi_a+\kappa_{abc}\(b^b+\I t^b \, \frac{1/t-t}{1/t+t}\) \delta\xi^c,
\\
\hat \delta_+ \alpha =&\, \delta\alpha+\tau\, \delta \txi_0
-\hf\,\kappa_{abc}\(b^a+\I t^a \, \frac{1/t-t}{1/t+t}\)
\(\xi^b-c^b+\tau b^b-\frac{2t^b(\xi^0-\tau)}{1/t+t}\)\delta\xi^c,
\\
\hat\delta_-\alpha =&\, \delta\alpha+\bar\tau\, \delta\txi_0
-\hf\,\kappa_{abc}\(b^a+\I t^a \, \frac{1/t-t}{1/t+t}\)
\(\xi^b-c^b+\bar\tau b^b+\frac{2t^b(\xi^0-\bar\tau)}{1/t+t}\) \delta\xi^c\ .
\end{split}
\ee
These combinations have been designed such that the complicated transformation rules
of $\delta\txi_\Lambda,\delta\alpha$, obtained by linearizing \eqref{SL2Zxi}, translate into
the simple properties
\be
\label{SL2hatdel}
\hat\delta\txi_a\ \mapsto \ \hat\delta \xi_a,
\qquad
\hat\delta_+ \alpha \ \mapsto \ \frac{\hat\delta_+\alpha}{c\tau+d},
\qquad
\hat\delta_- \alpha\ \mapsto \ \frac{\hat\delta_-\alpha}{c\bar\tau+d},
\ee
which we now need to prove for the instanton contributions given by \eqref{twlineIIBtxi}.
To this end, we note that the result \eqref{StransderG} implies that
$-\pa_{\txi_a}G, \pa_{\xi^a}G, (1-\xi^\Lambda\pa_{\xi^\Lambda})G$ transform in the same way as
$\delta\xi^a,\delta\txi_\Lambda,\delta\alpha$. Due to this, from these derivatives
one can form similar linear combinations
\bea
\hat\pa_{\xi^a} G &=&\pa_{\xi^a} G    - \kappa_{abc}\(b^b+\I t^b \, \frac{1/t-t}{1/t+t}\) \pa_{\txi_c} G,
\label{defdeltahG} \\
\hat G_+ &=&(1-\xi^\Lambda\pa_{\xi^\Lambda}+ \tau \pa_{\xi^0})G +\hf\,\kappa_{abc}\(b^a+\I t^a \,
\frac{1/t-t}{1/t+t}\)\(\xi^b-c^b+\tau b^b-\frac{2t^b(\xi^0-\tau)}{1/t+t}\) \pa_{\txi_c}G,
\nn\\
\hat G_- &=&(1-\xi^\Lambda\pa_{\xi^\Lambda}+ \bar\tau \pa_{\xi^0})G
+\hf\,\kappa_{abc}\(b^a+\I t^a \, \frac{1/t-t}{1/t+t}\)
\(\xi^b-c^b+\bar\tau b^b+\frac{2t^b(\xi^0-\bar\tau)}{1/t+t}\) \pa_{\txi_c}G,
\nn
\eea
which satisfy the same transformation laws as in \eqref{SL2hatdel},
\be
\hat\pa_{\xi^a} G\ \mapsto \ \hat\pa_{\xi^a} G,
\qquad
\hat G_+ \ \mapsto \ \frac{\hat G_+}{c\tau+d},
\qquad
\hat G_-\ \mapsto \ \frac{\hat G_-}{c\bar\tau+d}.
\ee
Using \eqref{twlineIIBtxi} to express \eqref{defdeltahx} in terms of \eqref{defdeltahG},
one finds
\bea
\label{defdeltahx2}
\hat\delta\txi_a &=&
- {\sum_{m,n}}' \oint_{C_{m,n}} \frac{\de \varpi'}{2 \pi \I \varpi'} \,
\left[ K(\varpi,\varpi')\, \hat\pa_{\xi^a} G
-2\I\, \frac{\kappa_{abc}t^b\p_{\txi_c}G_{m,n}}{\(1/t'+t'\)^2} \right],
\\
\hat\delta_+\alpha&=&
- {\sum_{m,n}}' \oint_{C_{m,n}} \frac{\de \varpi'}{2 \pi \I \varpi'} \,
\left[ K(\varpi,\varpi')\,\hat G_+
+2\I\kappa_{abc}\(\frac{(1+tt')\tau_2 t^a}{(t-\I)(t'-\I)}- (c^a-\tau b^a)\)
\frac{t^b\p_{\txi_c}G_{m,n}}{(1/t'+t')^2}\],
\nn
\\
\hat\delta_-\alpha& =&- {\sum_{m,n}}' \oint_{C_{m,n}} \frac{\de \varpi'}{2 \pi \I \varpi'} \,
\left[ K(\varpi,\varpi')\,\hat G_-
-2\I\kappa_{abc}\(\frac{(1+tt')\tau_2 t^a}{(t+\I)(t'+\I)}+(c^a-\bar\tau b^a)\)
\frac{t^b\p_{\txi_c}G_{m,n}}{(1/t'+t')^2}\].
\nn
\eea
In each of these expressions, the first term in the square bracket manifestly
transforms as in \eqref{SL2hatdel}. To see that this is also true for the remaining terms,
we rewrite them in terms of the Cayley-rotated coordinate $z$ as
\be
\begin{split}
\hat\delta^{(2)}\txi_a =&\,
- \kappa_{abc}t^b {\sum_{m,n}}' \oint_{C_{m,n}} \frac{\de z'}{8\pi^2 z'^2}\,
(1-z'^2)\p_{\txii{0}_c}G_{m,n},
\\
\hat\delta^{(2)}_+\alpha=&\,-
\kappa_{abc} t^a {\sum_{m,n}}' \oint_{C_{m,n}}  \frac{\de z'}{8\pi z'^2}
\left( c^b - \tau b^b -  \frac12 (z+z') \tau_2 t^b \right) (1-z'^2)\p_{\txii{0}_c}G_{m,n},
\\
\hat\delta^{(2)}_-\alpha=&\, \kappa_{abc} t^a {\sum_{m,n}}' \oint_{C_{m,n}}  \frac{\de z'}{8\pi z'^2}
\left( c^b -\bar \tau b^b + \frac12 \frac{z+z'}{z z'} \tau_2 t^b \right) (1-z'^2)\p_{\txii{0}_c}G_{m,n}.
\end{split}
\ee
Given that, under S-duality,
\be
(1-z^2)\pa_{\txi_a} G\ \mapsto \ \frac{(1-z^2)\pa_{\txi_a} G}{c\tau+d},
\ee
our claim trivially follows. It is interesting to note that the contribution
$\hat\delta^{(2)}\txi_a$ can be removed by  adjusting the mirror map
\eqref{mirror-map} by a further shift of $\tzeta_a$ by
\be
\Delta\tilde\zeta_a = -\kappa_{abc}t^b {\sum_{m,n}}' \oint_{C_{m,n}} \frac{\de \varpi}{\pi \varpi} \,
\frac{\p_{\txii{0}_c}G_{m,n}}{\(1/t+t\)^2}.
\label{addtzeta}
\ee
However, such a correction is modular invariant by itself and thus
represents an inherent ambiguity in the definition of the mirror-map.

Finally, we turn to the contact potential \eqref{fullephi}. In the one-instanton approximation,
taking into account the correction to $Y^a$ coming from \eqref{mirror-map}, we find
\be
\label{contrPhi}
e^\Phi = \frac{\tau_2^2}{2}\, \cV(t^a)
- \frac{1}{16\pi} {\sum_{m,n}}' \oint_{C_{m,n}} \frac{\de \varpi}{\varpi}\, A,
\ee
where
\be
\begin{split}
A=&\,  \(\varpi^{-1} Y^{\Lambda}-\varpi \bar Y^{\Lambda} \)\p_{\xi^\Lambda}G_{m,n}
+\(\varpi^{-1} F_{a}(Y)-\varpi \bar F_{a}(\bar Y) \)\p_{\txi_a}G_{m,n}
\\
&\,
+2\tau_2 \kappa_{abc} t^a t^b \frac{t^{-1}-t}{(t^{-1}+t)^2}\p_{\txi_c}G_{m,n}.
\end{split}
\ee
Using the transformation properties
\be
\begin{split}
&
\frac{\de \varpi}{\varpi}\ \mapsto\ \frac{|c\tau+d|}{c\xi^0+d}\,\frac{\de \varpi}{\varpi} ,
\qquad
\frac{\varpi^{-1}-\varpi }{(\varpi^{-1}+\varpi)^2}
\ \mapsto\
\frac{(c\xi^0+d)^2}{|c\tau+d|^2}\,\frac{\varpi^{-1}-\varpi }{(\varpi^{-1}+\varpi)^2}
-\frac{c\tau_2}{2}  \frac{c\xi^0+d}{|c\tau+d|^2} ,
\\
&\qquad\quad
A \ \mapsto\
\frac{(c\xi^0+d)}{|c\tau+d|^2}\,A
+\frac{c (c\xi^0+d)}{|c\tau+d|^2}\,
t \p_t\, \(\frac{t^{-1}+t}{c\xi^0+d}\, G_{m',n'}\)
+\mbox{reg.}
\end{split}
\label{transX}
\ee
we see that the integrand transforms by a total derivative,
and therefore that the contact potential transforms as
in \eqref{SL2phi}.

\providecommand{\href}[2]{#2}\begingroup\raggedright\endgroup


\end{document}